\documentclass[english,aps,prc,nofootinbib,superscriptaddress,twocolumn,letter]{revtex4}
\usepackage[latin9]{inputenc}
\usepackage{hyperref}
\hypersetup{
  colorlinks=true,        % false: boxed links; true: colored links
  linkcolor=blue,         % color of internal links
  citecolor=cyan,         % color of links to bibliography
}
\usepackage{breakurl}
\usepackage{graphicx}
\usepackage{epsf}
\usepackage{epsfig}
\usepackage{amssymb,amsmath}
\usepackage[usenames]{color}
\usepackage{amssymb}
\usepackage{times}
\usepackage{comment}

\allowdisplaybreaks



\begin{document}

\preprint{This line only printed with preprint option}

\title{1+1 dimensional relativistic viscous non-resistive magnetohydrodynamics with longitudinal boost invariance}

\author{Ze-Fang Jiang}
% \email{jiangzf@mails.ccnu.edu.cn}
\affiliation{Department of Physics and Electronic-Information Engineering, Hubei Engineering University, Xiaogan, Hubei, 432000, China}
\affiliation{Institute of Particle Physics and Key Laboratory of Quark and Lepton Physics (MOE), Central China Normal University, Wuhan, Hubei, 430079, China}

\author{Shuo-Yan Liu}
\affiliation{Department of Physics and Electronic-Information Engineering, Hubei Engineering University, Xiaogan, Hubei, 432000, China}

\author{Tian-Yu Hu}
\affiliation{Department of Physics and Electronic-Information Engineering, Hubei Engineering University, Xiaogan, Hubei, 432000, China}

\author{Huang-Jing Zheng}
\affiliation{Department of Science, Three Gorges University, Yichang, Hubei, 443002, China}

\author{Duan She}
\email{sheduan@hnas.ac.cn}
\affiliation{Institute of Physics, Henan Academy of Sciences, Zhengzhou 450046, China}

\begin{abstract}
We study 1+1 dimensional relativistic non-resistive magnetohydrodynamics (MHD) with longitudinal boost invariance and shear stress tensor. Several analytical solutions that describe the fluid temperature evolution under the equation of state (EoS) $\varepsilon=3p$ are derived, relevant to relativistic heavy-ion collisions. Extending the Victor-Bjorken ideal MHD flow to include non-zero shear viscosity, two perturbative analytical solutions for the first-order (Navier-Stokes) approximation are obtained. For small, power-law evolving external magnetic fields, our solutions are stable and show that both magnetic field and shear viscosity cause fluid heating with an early temperature peak, align with the numerical results. In the second-order (Israel-Stewart) theory, our findings show that the combined presence of magnetic field and shear viscosity leads to a slow cooling rate of fluid temperature, with initial shear stress significantly affecting temperature evolution of QGP.
\end{abstract}


\maketitle

\section{Introduction}

Heavy-ion collisions offer a unique opportunity to explore the properties of hot, dense nuclear matter - quark-guluon plasma (QGP)- created at RHIC and LHC~\cite{STAR:2005gfr,ALICE:2008ngc}. Recently research has revealed that these collisions generate extremely strong magnetic fields, ranging from 10$^{18}$ to 10$^{19}$ Gauss, due to the rapid motion of positively charged spectators~\cite{Deng:2012pc,Li:2016tel,Gursoy:2014aka}. These huge magnetic fields are expected to profoundly impact the dynamics of the QGP~\cite{Gursoy:2014aka,Das:2016cwd,Gursoy:2018yai,Chatterjee:2018lsx,Oliva:2020doe,Sun:2020wkg,Jiang:2022uoe}. The combination of huge magnetic fields and quantum anomalies can induce special transport phenomena known as anomalous transports. One such phenomenon is the chiral magnetic effect (CME)~\cite{Kharzeev:2007jp,Fukushima:2008xe,Kharzeev:2015znc,Kharzeev:2024zzm}, which predicts charge separation in a chirality-imbalanced medium. Additionally, a chiral current can be induced by the magnetic field, leading to the chiral separation effect (CSE)~\cite{Huang:2013iia,Pu:2014cwa}. Together, these effects can expected to induce a density wave called the ``chiral magnetic wave'' (CMW), which might disrupt the elliptic flow degeneracy between $\pi^{\pm}$~\cite{Kharzeev:2010gd}. Over the past decade, experimental collaborations at RHIC and LHC have made significant efforts to observe signals of the CME, CSE, and CMW, attracting considerable attention in studying hot, dense nuclear matter under strong magnetic fields~\cite{STAR:2015wza,CMS:2017lrw,STAR:2021mii,STAR:2021pwb,STAR:2019clv,ALICE:2019sgg,STAR:2023jdd}. However, it is still a challenge to extract the signals from the huge backgrounds caused by the collective flows.

The evolution of the QGP, mainly driven by pressure gradients, has been well-described by relativistic hydrodynamic, explaining experimentally measured harmonics flow and global polarization in non-central nucleus-nucleus collisions~\cite{Song:2010mg,Heinz:2013th,Gale:2013da,Karpenko:2013wva,Jeon:2015dfa,Becattini:2017gcx,Becattini:2020ngo,Csanad:2016add,Ze-Fang:2017ppe,Zhao:2022ayk,Zhao:2020wcd,Jiang:2021ajc,Wu:2021fjf,Jiang:2024ekh}.
Fully understand the effects of strong magnetic fields on QGP, solving the (3+1)-dimensional relativistic magnetohydrodynamics (MHD) equations is necessary~\cite{Inghirami:2016iru,Nakamura:2022idq,Mayer:2024kkv}, as they account for the magnetic field's dynamic coupling with the QGP fluid.
%While the magnetic field generated in heavy-ion collisions rapidly decays in the vacuum~\cite{}, the conducting plasma created in these collisions can substantially delay this decay through the generation of induction currents due to Lenz's law.
Lattice-QCD calculation indicates that the QGP has a finite temperature-dependent electrical conductivity ($\sigma_{el}$)~\cite{Ding:2010ga,Ding:2016hua}. However, the interaction of the initial magnetic field with the QGP, as well as its subsequent evolution, are still not fully understood and are currently under active investigation~\cite{Bali:2013owa,Pang:2016yuh,Jiang:2024bez,Huang:2024aob}. The relative significance of an external magnetic field on fluid evolution can be gauged using the dimensionless ratio $\sigma \equiv B^{2}/T^{4}$, which compares magnetic-field energy density to QGP temperature. When $\sigma>1$, one found that the magnetic field's influence on fluid evolution becomes notable and must be considered~\cite{Roy:2015kma,Pu:2016ayh,Siddique:2019gqh}.

In recent works, the influence of electromagnetic fields on the quark-gluon plasma (QGP) fluid in special relativistic systems has been investigated under hydrodynamic frame~\cite{Fukushima:2024tkz}. Prior research studied the 1+1 dimensional flow, utilizing the longitudinally boost-invariant Bjorken flow model within the framework of MHD without dissipative effect~\cite{Roy:2015kma,Pu:2016ayh,Siddique:2019gqh,Peng:2022cya}. These studies incorporated a transverse and time-dependent homogeneous magnetic field while neglecting dissipative effects such as viscosity and thermal conduction. Notably, in ideal MHD, the evolution of energy density maintains the same decay rate as Bjorken flow, attributed to the ``frozen-flux theorem"~\cite{Roy:2015kma}. Subsequently, nonzero magnetization was introduced into the MHD~\cite{Pu:2016ayh}, the longitudinal expansion effect was examined~\cite{She:2019wdt,HaddadiMoghaddam:2020ihi}, and self-similar rotating solutions were derived within the context of MHD~\cite{Shokri:2018qcu}.

Various studies have also aimed to analyze the stability and causality of relativistic dissipative fluid dynamics within the frameworks of both standard and modified Israel-Stewart (IS) theories, in the presence of a magnetic field~\cite{Biswas:2020rps,Cordeiro:2023ljz,Fang:2024skm,Rocha:2023ilf,Most:2021rhr}.
It has been found that a straightforward extension of non-relativistic viscous fluid formulations (also known as Navier-Stokes theory) to the relativistic regime leads to acausal and linearly unstable behavior. Subsequently, these issues were addressed using the IS theory to develop a causal and stable second-order formalism. These studies have systematically enhanced understanding of relativistic non-resistive magnetohydrodynamics (MHD).

In this paper, we extend previous studies by incorporating shear viscosity into the MHD equations within the Azwinndini-Bjorken viscous flow framework~\cite{Muronga:2001zk,Muronga:2003ta}, and solve them both analytically and numerically. Starting with the 1+1 dimensional relativistic non-resistive MHD \cite{Biswas:2020rps}, which includes shear viscosity and a magnetic field, we derive a series of novel solutions for non-resistive MHD. Our results indicate that for small, power-law evolving magnetic fields, our analytical solutions are stable and reveal that both the magnetic field and shear viscosity contribute to fluid heating, with an early temperature peak that aligns with numerical findings. In the context of the second-order (Israel-Stewart) theory, we find that the combined presence of a magnetic field and shear viscosity results in a slower cooling rate of fluid temperature. Furthermore, we note that our assumptions and solutions are not only straightforward but also readily adaptable to other MHD studies.

This paper is organized as follows. In Sec.~\ref{sect:eqtherm}, we introduce the MHD framework with dissipative effects. In Sec.~\ref{section-3}, we present the solutions for the MHD with first-order (NS) theory. In Sec.~\ref{section-4}, we show the results for the MHD in the present of second-order (IS) theory. Finally, we summarize and conclude in the Sec.~\ref{section-5}. Throughout this work, $u^{\mu}=\gamma\left(1,\overrightarrow{\boldsymbol{v}}\right)$ is the four-velocity field that satisfies $u^{\mu}u_{\mu}=1$ and the spatial projection operator $\Delta^{\mu\nu}=g^{\mu\nu}-u^{\mu}u^{\nu}$ is defined with the Minkowski metric $g^{\mu\nu}={\rm diag}\left(1,-1,-1,-1\right)$. It is note-worthy that the orthogonality relation $\Delta^{\mu\nu}u_{\nu}=0$ is satisfied.

\section{Relativistic viscous non-resistive magnetohydrodynamic}
\label{sect:eqtherm}

In this work, we consider the causal relativistic second order theory for relativistic fluids by Israel-Stewart (IS) in presence of a magnetic field given in Refs.~\cite{Denicol:2018rbw,Biswas:2020rps}. The total energy-momentum tensor of the viscous fluid can be written as
\begin{equation}
\begin{aligned}
T^{\mu\nu} &= (\varepsilon + p + \Pi + E^{2}+ B^{2})u^{\mu}u^{\nu} \\
           &-\left(p+ \Pi+\frac{1}{2}E^{2}+\frac{1}{2}B^{2}\right)g^{\mu\nu}\\
&-E^{\mu}E^{\nu}-B^{\mu}B^{\nu}-u^{\mu}\epsilon^{\nu\lambda\alpha\beta}E_{\lambda}B_{\alpha}u_{\beta} \\
& -u^{\nu}\epsilon^{\mu\lambda\alpha\beta}E_{\lambda}B_{\alpha}u_{\beta} + \pi^{\mu\nu},
\label{tmunu_total}
\end{aligned}
\end{equation}
where $\varepsilon$, $p$ are energy density, pressure, $u^{\mu}$ is the four velocity field, and $\pi^{\mu\nu}$, $\Pi$ are shear viscous tensor and bulk viscous pressure. The magnetic field and electric field four vectors are
\begin{equation}
\begin{aligned}
B^{\mu} = \frac{1}{2}\epsilon^{\mu\nu\alpha\beta}u_{\nu}F_{\alpha\beta},~~E^{\mu}=F^{\mu\nu}u_{\nu},
\end{aligned}
\end{equation}
which satisfies $u^{\mu}E_{\mu}=0$ and $u^{\mu}B_{\mu}=0$ meaning that both $E^{\mu}$ and $B^{\mu}$ are spacelike. The modulus $B^{\mu}B_{\mu}=-B^{2}$ and $E^{\mu}E_{\mu}=-E^{2}$. Here, $\epsilon^{\mu\nu\alpha\beta}$ is the Levi-Civita tensor satisfying $\epsilon^{0123}=-\epsilon_{0123}=+1$, $F^{\mu\nu}$ is the Faraday tensor satisfying $F^{\mu\nu}=(\partial^{\mu}A^{\nu}-\partial^{\nu}A^{\mu})$.
The non-resistance limit means the electric conductivity $\sigma_{el}$ is infinite. In this limit, in order to keep the electric charge current $j^{\mu}=\sigma_{el}E^{\mu}$ be finite, we assume the $E^{\mu}\rightarrow 0$. Then the relevant Maxwell's equations which govern the evolution of magnetic fields in the fluid is $\partial_{\nu}(B^{\mu}u^{\nu}-B^{\nu}u^{\mu})=0$. Thus the energy-momentum tensor reduces to~\cite{Roy:2015kma,Biswas:2020rps}
\begin{equation}
\begin{aligned}
T^{\mu\nu} &= (\varepsilon + p + \Pi + B^{2})u^{\mu}u^{\nu}-\left(p+\Pi+\frac{1}{2}B^{2}\right)g^{\mu\nu} \\
           &~~~~-B^{\mu}B^{\nu}+\pi^{\mu\nu}.
\label{tmunu_t}
\end{aligned}
\end{equation}

The space-time evolution of the fluid and electric-magnetic fields are described by the energy-momentum conservation
\begin{equation}
\begin{aligned}
\partial_{\mu}T^{\mu\nu}=0.
\end{aligned}
\end{equation}
To close the system of equations, we need to choose the Equation of State (EoS) for the thermodynamic quantities. In the hot limit with high temperature, we consider a conformal fluid such that $\varepsilon=3 p$, where $c_{s}^{2}=1/3$.
%We note in a fully realistic solution, one should use the lattice QCD equation of state, with the speed of sound being a temperature dependent function. However, in current work we approximate $c_s(T)$ with a temperature independent constant $c_s$.
For simplicity, we also neglect the magnetization of the QGP~\cite{Pu:2016ayh}, which implies an isotropic pressure and no change in the Equation of State (EoS) of the fluid due to magnetic field.

Since we assume longitudinal boost invariance for the fluid, it is convenient to introduce Milne coordinates, defined as $t=\tau \cosh\eta$ and $z=\tau\sinh\eta$, where $\tau=\sqrt{t^{2}-z^{2}}$ being the proper time, and $\eta=0.5\ln[(t+z)/(t-z)]$ being the space-time rapidity. The fluid velocity can be written as
\begin{equation}
\begin{aligned}
u^{\mu} = (\cosh\eta, 0, 0, \sinh\eta)=\gamma(1,0,0,z/t),
\end{aligned}
\end{equation}
where $\gamma=\cosh\eta$ is the Lorentz contraction factor.

Following the refs.~\cite{Roy:2015kma,Pu:2016ayh,She:2019wdt,Biswas:2020rps}, we consider a simple homogeneous magnetic field obeys a power-law decay in proper time as follow,
\begin{equation}
\overrightarrow{B}(\tau)=\overrightarrow{B}_{0}\left(\frac{\tau_{0}}{\tau}\right)^{a},
\label{11}
\end{equation}
where $a$ is a decay constant and $a>0$. $\tau_0$ is the initial proper time and $B_{0}\equiv B(\tau_{0})$ is the initial strength of magnetic field.

The energy conservation equation is derived by projecting the energy-momentum tensor conservation law $\partial_{\mu}T^{\mu\nu}=0$ along the fluid four-velocity $u_\mu$, in the Landau-Lifshitz frame, one has
\begin{equation}
\begin{aligned}
u_{\mu}\partial_{\nu}T^{\mu\nu} &= D\left(\varepsilon + \frac{1}{2}B^{2}\right)  + (\varepsilon+p+B^{2})\theta - \pi^{\mu\nu}\sigma_{\mu\nu} - \Pi\theta\\
&=\frac{\partial \left(\varepsilon + \frac{1}{2}B^{2}\right)}{\partial \tau}  + \frac{\varepsilon+p+B^{2}}{\tau} - \frac{1}{\tau}\pi - \Pi\frac{1}{\tau} \\
&=0,
\label{energy_conservation_1}
\end{aligned}
\end{equation}
where $D=u\cdot \partial=\partial/\partial\tau$, $\theta\equiv\partial_{\mu}u^{\mu}=\nabla_{\mu}u^{\mu}=\frac{1}{\tau}$ is the expansion factor and $\pi=\pi^{00}-\pi^{33}$~\cite{Muronga:2001zk,Muronga:2003ta}.

Similarly, the projection of the energy-momentum equation onto the direction orthogonal to $u^{\mu}$,
\begin{equation}
\begin{aligned}
(g_{\mu\nu}-u_{\mu}u_{\nu})\partial_{\alpha}T^{\alpha\nu} = 0,
\end{aligned}
\end{equation}
leads to the momentum-conservation equation
\begin{equation}
\begin{aligned}
\nabla^{\mu}\left(p+\Pi+\frac{B^{2}}{2}\right) & -(\varepsilon+p+\Pi+B^{2})\frac{\partial u^{\mu}}{\partial\tau}\\
 & -\Delta_{\nu}^{\mu}\nabla_{\sigma}\pi^{\nu\sigma}+\pi^{\mu\nu}\frac{\partial u_{\nu}}{\partial\tau}=0,\label{euler}
\end{aligned}
\end{equation}
where $\nabla^{\mu}=\Delta^{\mu\nu}\partial_{\nu}$ is the gradient operator. The last three terms vanish due to the longitudinal boost invariance and $u^{x}=u^{y}=0$ in transverse direction~\cite{Muronga:2003ta,Roy:2015kma}. Note that for $\mu=\eta$, it reads
\begin{equation}
\begin{aligned}
\frac{\partial}{\partial\eta}\left(p+\Pi+\frac{1}{2} B^{2}\right)=0,
\end{aligned}
\end{equation}
thus showing that all thermodynamical variables depend only on $\tau$ and are otherwise uniform in space. Since the velocities in the $x-$ an $y-$directions are initially zero, they will remain so also at later times.

Since we fouce on the non-resistive viscous hydrodynamic, in the second order Israel-Stewart (IS) theory, the viscous stresses $\Pi$, $\pi^{\mu\nu}$ are considered as independent dynamical variables and given by (e.g. see ref~\cite{Muronga:2003ta}),
\begin{equation}
\begin{aligned}
\frac{\partial\Pi}{\partial \tau} = -\frac{\Pi}{\tau_{\Pi}} - \frac{1}{2}\frac{1}{\beta_{0}}\Pi\left[\beta_{0}\frac{1}{\tau} + T\frac{\partial}{\partial\tau} \left(\frac{\beta_{0}}{T}\right)\right] - \frac{1}{\beta_{0}}\frac{1}{\tau},
\label{Pimunu_t}
\end{aligned}
\end{equation}

\begin{equation}
\begin{aligned}
\frac{\partial\pi^{\mu\nu}}{\partial \tau} = &-\frac{\pi^{\mu\nu}}{\tau_{\pi}} - \frac{1}{2}\frac{1}{\beta_{2}}\pi^{\mu\nu}\left[\beta_{2}\frac{1}{\tau} + T\frac{\partial}{\partial\tau} \left(\frac{\beta_{2}}{T}\right)\right] \\
&+ \frac{1}{\beta_{2}}\left[\widetilde{\Delta}^{\mu\nu}-\frac{1}{3}\Delta^{\mu\nu}\right]\frac{1}{\tau},
\label{pimunu_t_1}
\end{aligned}
\end{equation}
where the relaxation times $\tau_{\Pi}$ and $\tau_{\pi}$ are
\begin{equation}
\begin{aligned}
\tau_{\Pi} = \zeta\beta_{0},~~~~\tau_{\pi} = 2\eta \beta_{2}.
\label{rt_t}
\end{aligned}
\end{equation}
Here $\zeta$ and $\eta$ are the bulk and shear viscosity coefficients, respectively. $\beta_{0}$ and $\beta_{2}$ are the transport coefficients. In the final term of Eq.~(\ref{pimunu_t_1}), $\widetilde{\Delta}^{\mu\nu}=\Delta^{\mu\nu}$ for $0\leq\mu$, $\nu\leq$1, and is zero otherwise (due to the fact that there is only one non-vanishing spatial component in the four-velocity).

For the sake of convenience, some useful equations and relations of thermodynamic can be written as~\cite{Muronga:2003ta}:
\begin{equation}
\begin{aligned}
p&=a_{1}T^{4},~~p=c_{s}^{2}\varepsilon=\frac{1}{3}\varepsilon,~~\eta = b_{1}T^{3},\\
\zeta &= b_{2}T^{3}, ~~\frac{\eta}{s} = \frac{b_{1}}{4a_{1}},~~\frac{\zeta}{s} = \frac{b_{2}}{4a_{1}}, \\
B^{2}_{0}&= \sigma T^{4}_{0}, ~~
T(\tau_{0})=T_{0},~~\Tilde{T} = T/T_{0},~~\Tilde{T}(\tau_{0}) = 1.
\label{e_simp}
\end{aligned}
\end{equation}
where $\Tilde{T}$ is the normalized dimensionless temperature, while $T_{0}$ is the temperature at initial proper time $\tau_{0}$. We assume $T_{0}=T(\tau_{0})=0.65$ GeV, $\tau_{0}=0.6$ fm/c and
\begin{equation}
\begin{aligned}
a_{1}=\left(16+\frac{21}{2}N_{f}\right)\frac{\pi^{2}}{90}
\label{rt_t}
\end{aligned}
\end{equation}
is a constant determined by the number of quark flavors and the number of gluon colors~\cite{Muronga:2003ta}, $s$ is the entropy density, and
\begin{equation}
\begin{aligned}
b_{1}=(1+1.7N_{f})\frac{0.342}{(1+N_{f}/6)\alpha_{s}^{2}\ln(\alpha_{s}^{-1})}.
\label{rt_t}
\end{aligned}
\end{equation}
Here $N_{f}$ is the number of quark flavors, taken to be 3, and $\alpha_{s}$ is the strong fine structure constant, taken to be 0.4-0.5~\cite{Muronga:2003ta}.
In the context of a conformal fluid (or a system of massless particles), the bulk viscosity $\zeta=0$ since the bulk viscosity does not apply to such systems~\cite{Muronga:2003ta}.

\section{Analytic solutions}
\label{section-3}
%\subsection{Bjorken flow}

%For the magnetic field is zero, and shear and bulk viscosity is zero, the Equation of State refers to the case $c_{s}=1\sqrt{3}$ and $\varepsilon=3p$, then the Eq.~\ref{energy} becomes
%\begin{equation}
%\begin{aligned}
%\frac{\partial \varepsilon}{\partial\tau}  + \frac{\varepsilon+p}{\tau} = 0, \\
%\Rightarrow \frac{\partial (3a_{1}T^{4})}{\partial\tau}  + \frac{4a_{1}T^{4}}{\tau} = 0, \\
%\Rightarrow 12 a_{1}T^{3}\cdot \frac{\partial T}{\partial\tau}  + \frac{4a_{1}T^{4}}{\tau} = 0, \\
%\Rightarrow \frac{\partial \Tilde{T}}{\partial\tau} + \frac{\Tilde{T}}{3\tau} = 0.\\
%\end{aligned}
%\end{equation}
%The analytic solution of this equation is
%\begin{equation}
%\begin{aligned}
%\Tilde{T} = \frac{T}{T_{0}} = \left( \frac{\tau_{0}}{\tau}\right)^{\frac{1}{3}}.
%\end{aligned}
%\end{equation}

\subsection{Analytical solution for the ideal MHD}
\label{section-2-a}

\begin{figure}[tbp!]
\includegraphics[width=0.85\linewidth]{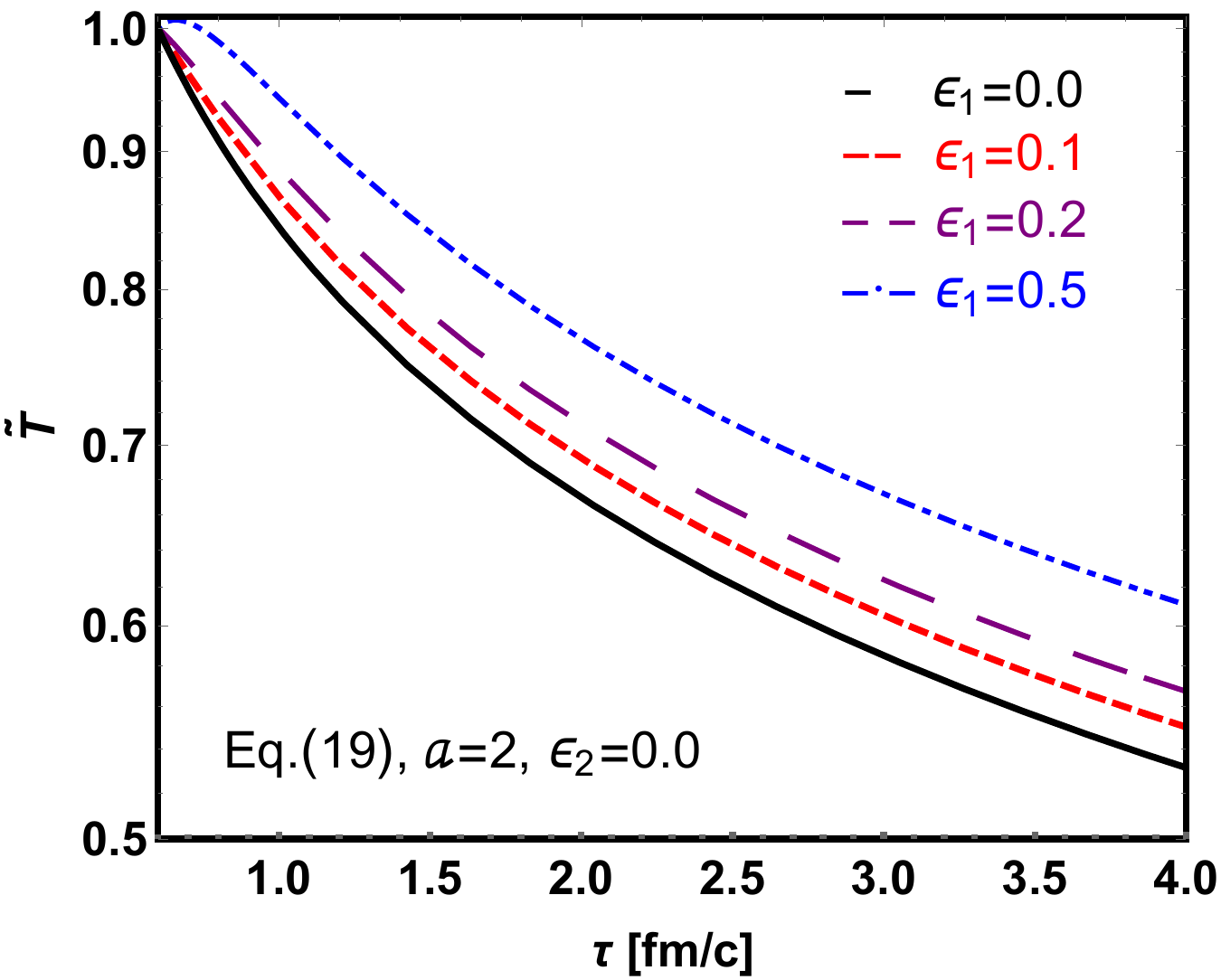} \\
\includegraphics[width=0.85\linewidth]{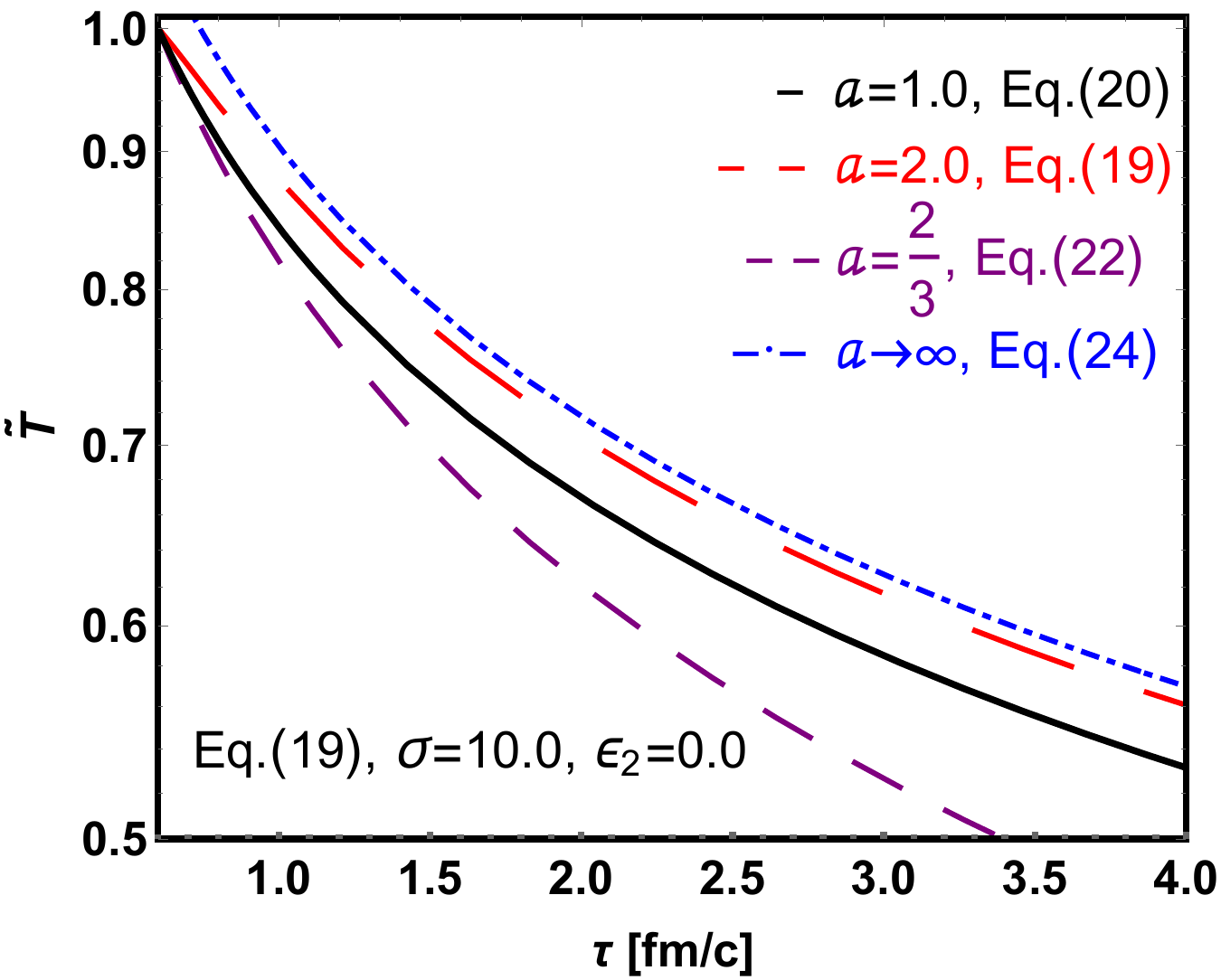}
\caption{(Color online) Evolution of normalized temperature $\Tilde{T}=T/T_{0}$ for a magnetic field with as functions of proper time $\tau$ for different initial magnetic field (upper panel) and different magnetic field decay parameter $a$ (lower panel).}
\label{f:ideal-mhd}
\end{figure}

Let's begin by addressing the 1+1 dimensional ideal MHD flow (\textit{Victor-Bjorken} flow~\cite{Roy:2015kma}), starting from Eq.~(\ref{energy_conservation_1}). By neglecting the contribution from shear viscosity, the energy conservation equation can be written as
\begin{equation}
\begin{aligned}
\frac{\partial \left(\varepsilon + \frac{1}{2}B^{2}\right)}{\partial \tau}  + \frac{\varepsilon+p+B^{2}}{\tau} &= 0,\\
%\Rightarrow \frac{\partial \left(3a_{1}T^{4} + \frac{1}{2} B^{2}_{0}\left(\frac{\tau_{0}}{\tau}\right)^{2a}\right)}{\partial \tau} + \frac{4a_{1}T^{4}+B^{2}_{0}\left(\frac{\tau_{0}}{\tau}\right)^{2a}}{\tau} &= 0, \\
%\Rightarrow 12a_{1} T^{3}\frac{\partial T}{\partial \tau} - a B_{0}^{2}\left(\frac{\tau_{0}}{\tau}\right)^{2a-1}\frac{\tau_{0}}{\tau}\frac{1}{\tau} + \frac{4a_{1}T^{4}+B^{2}_{0}\left(\frac{\tau_{0}}{\tau}\right)^{2a}}{\tau} &= 0,\\
%\Rightarrow \frac{\partial T}{\partial \tau} + \frac{(1-a) B_{0}^{2}}{12 a_{1} T^{3}}\left(\frac{\tau_{0}}{\tau}\right)^{2a}\frac{1}{\tau} + \frac{T}{3\tau} &= 0,\\
\Rightarrow \frac{\partial T}{\partial \tau} + \frac{(1-a) \sigma T_{0}^{4}}{12 a_{1} T^{3}}\left(\frac{\tau_{0}}{\tau}\right)^{2a}\frac{1}{\tau} + \frac{T}{3\tau} &= 0,\\
\Rightarrow \frac{\partial \Tilde{T}}{\partial \tau} + \frac{(1-a) \sigma }{12 a_{1} \tilde{T}^{3}}\left(\frac{\tau_{0}}{\tau}\right)^{2a}\frac{1}{\tau} + \frac{\tilde{T}}{3\tau} &= 0.
\label{MHD_energy_1}
\end{aligned}
\end{equation}
For convenience, we introduce the parameter $\epsilon_{1}=\frac{(a-1)\sigma}{12 a_{1}}$, which is related to the strength parameter $\sigma$ and the decay rate $a$ of the magnetic field. With this definition, the energy conservation equation Eq.~(\ref{MHD_energy_1}) simplified to
\begin{equation}
\begin{aligned}
\frac{\partial \Tilde{T}}{\partial \tau} + \frac{\tilde{T}}{3\tau} - \frac{\epsilon_1}{\tilde{T}^{3}}\left(\frac{\tau_{0}}{\tau}\right)^{2a}\frac{1}{\tau} = 0.
\label{MHD_energy_2}
\end{aligned}
\end{equation}
The solution of Eq.~(\ref{MHD_energy_2}) is
\begin{equation}
\begin{aligned}
\Tilde{T}=\left(\frac{\tau_{0}}{\tau}\right)^{\frac{1}{3}}\left[1+\frac{6\epsilon_{1}}{3a-2}\left(1-\left(\frac{\tau_{0}}{\tau}\right)^{2a-\frac{4}{3}}\right)\right]^{\frac{1}{4}}.
\label{T_mhd_3}
\end{aligned}
\end{equation}

In the upper panel of Figure~\ref{f:ideal-mhd}, we plot the normalized dimensionless temperature $\Tilde{T}$ as a function of proper time $\tau$ for the case where $a=2$, considering various magnetic field parameter $\epsilon_{1}$. The initial proper time $\tau_{0}=0.6$ fm/c.
We find that for positive $\epsilon_{1}$ (as $a>1$), the normalized temperature $\Tilde{T}$ decays more slowly as $\epsilon_{1}$ increases. This is because the energy density is ``heated up'' by the rapid decay of the magnetic field~\cite{Roy:2015kma}. Furthermore, a larger initial magnetic field (determined by $\sigma$) leads to a slower decay of $\Tilde{T}$.

We observe that Eq.~(\ref{T_mhd_3}) has divergent behavior at different values of $a$, which determine the strength of the magnetic field. To illustrate this, we present following solutions under specific limits.

\emph{\textbf{Case-A}} For $a = 1$, thus $\epsilon_1\equiv 0$, which indicates the limit of infinite conductivity (or, the ideal MHD limit), and thus of a maximal magnetic induction. We obtain a solution of the ideal MHD, the solution type is as same as the Victor-Bjroken flow~\cite{Roy:2015kma},
\begin{equation}
\begin{aligned}
\Tilde{T} = \left(\frac{\tau_0}{\tau}\right)^{\frac{1}{3}}.
\end{aligned}
\end{equation}

\emph{\textbf{Case-B}} In the limit $a \rightarrow \frac{2}{3}$, one finds
\begin{equation}
\begin{aligned}
\lim\limits_{a \rightarrow \frac{2}{3}}\frac{6 \epsilon_1}{3a - 2} \left(1 - \left(\frac{\tau_0}{\tau}\right)^{2a - \frac{4}{3}}\right)
=\frac{\sigma}{9 a_{1}}\ln\left(\frac{\tau_{0}}{\tau}\right).
\label{eq:a32}
\end{aligned}
\end{equation}
After collecting terms, the solution for $a \rightarrow \frac{2}{3}$ is
\begin{equation}
\begin{aligned}
\Tilde{T}=\left(\frac{\tau_{0}}{\tau}\right)^{\frac{1}{3}}\left[1+\frac{\sigma}{9a_{1}}\ln\left(\frac{\tau_{0}}{\tau}\right)\right]^{\frac{1}{4}}.
\end{aligned}
\end{equation}
Note that for $\tau>\tau_{0}$, the log term is negative, hence reduces the value of $\Tilde{T}$, leading to a faster decrease of the temperature (this is shown in Fig.~\ref{f:ideal-mhd} lower panel magenta line).

\emph{\textbf{Case-C}} In the limit $a \rightarrow \infty$ and $\tau>\tau_{0}$, it is not difficult to find
\begin{equation}
\begin{aligned}
\lim\limits _{a\rightarrow\infty}\frac{6\epsilon_{1}}{3a-2}\left(1-\left(\frac{\tau_{0}}{\tau}\right)^{2a-\frac{4}{3}}\right)=\frac{\sigma}{6a_{1}}.
\end{aligned}
\end{equation}
Thus, this limit result in a solution as follow,
\begin{equation}
\begin{aligned}
\Tilde{T} = \left(\frac{\tau_0}{\tau}\right)^{\frac{1}{3}} \left( 1 + \frac{\sigma}{6 a_{1}}\right)^{\frac{1}{4}}.
\end{aligned}
\end{equation}
One finds a super-fast decay of the magnetic field (as $a \rightarrow \infty$) results to a slow decay of the temperature and poses a upper limit (this is shown in Fig.~\ref{f:ideal-mhd} lower panel purple line).

\subsection{Analytical solution for Navier-Stokes approximation viscous hydrodynamic}
\label{section-2-b}

\begin{figure}[tbp!]
\includegraphics[width=0.85\linewidth]{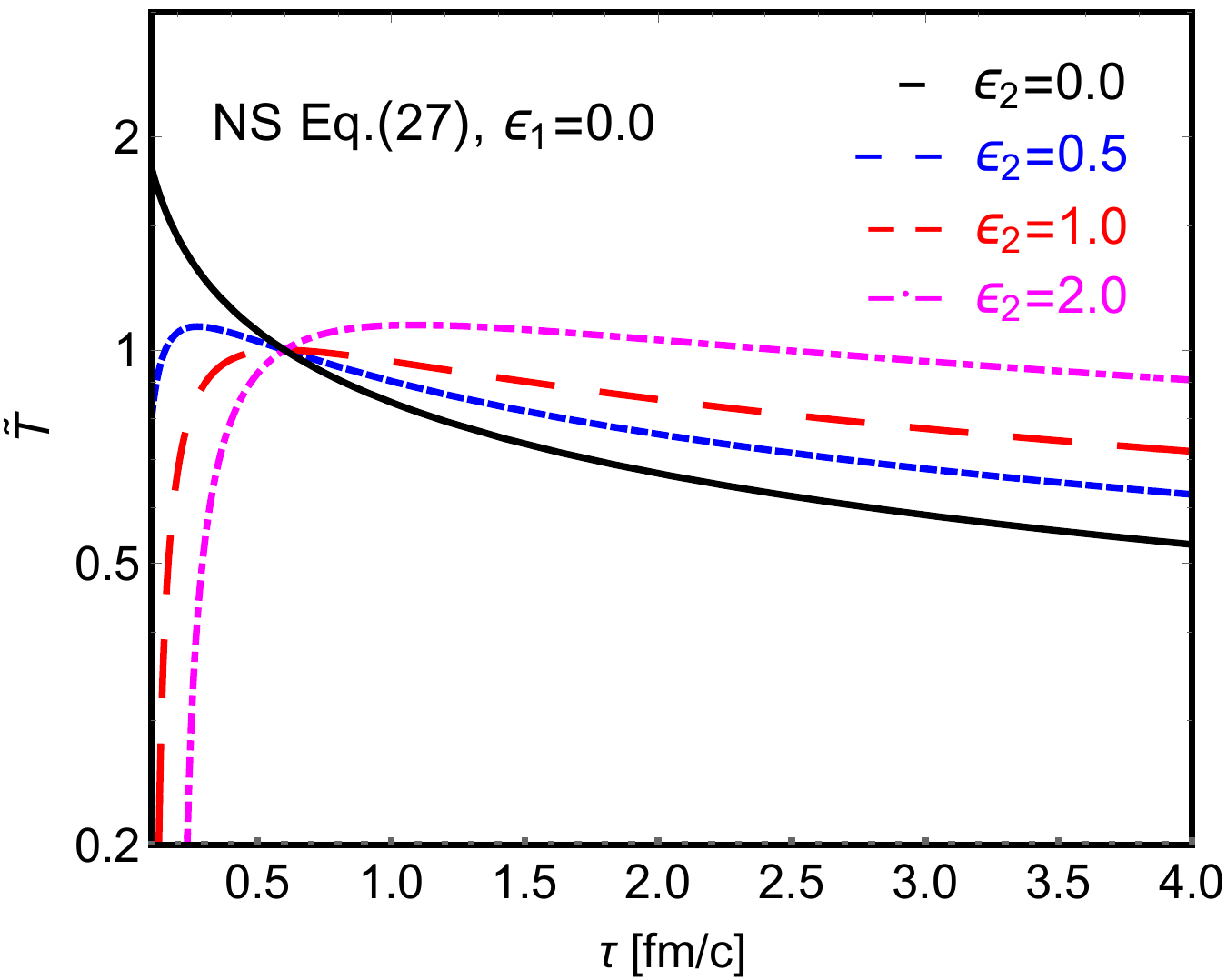}
\caption{(Color online) Evolution of normalized temperature $\Tilde{T}=T/T_{0}$ as functions of proper time $\tau$ for different shear viscosity value $\epsilon_{2}$, where $\epsilon_{2}=\frac{1}{9T_{0}}\frac{\eta}{s}$, $\tau_{0}=0.6$ fm/c, and $T_{0}=0.65$ GeV.}
\label{f:ns-1}
\end{figure}

%%%%%%%%%%%%%%%%%%%%%%%%%% Figure %%%%%%%%%%%%%%%%%%%%%%%%
\begin{figure}[tbp!]
\includegraphics[width=0.85\linewidth]{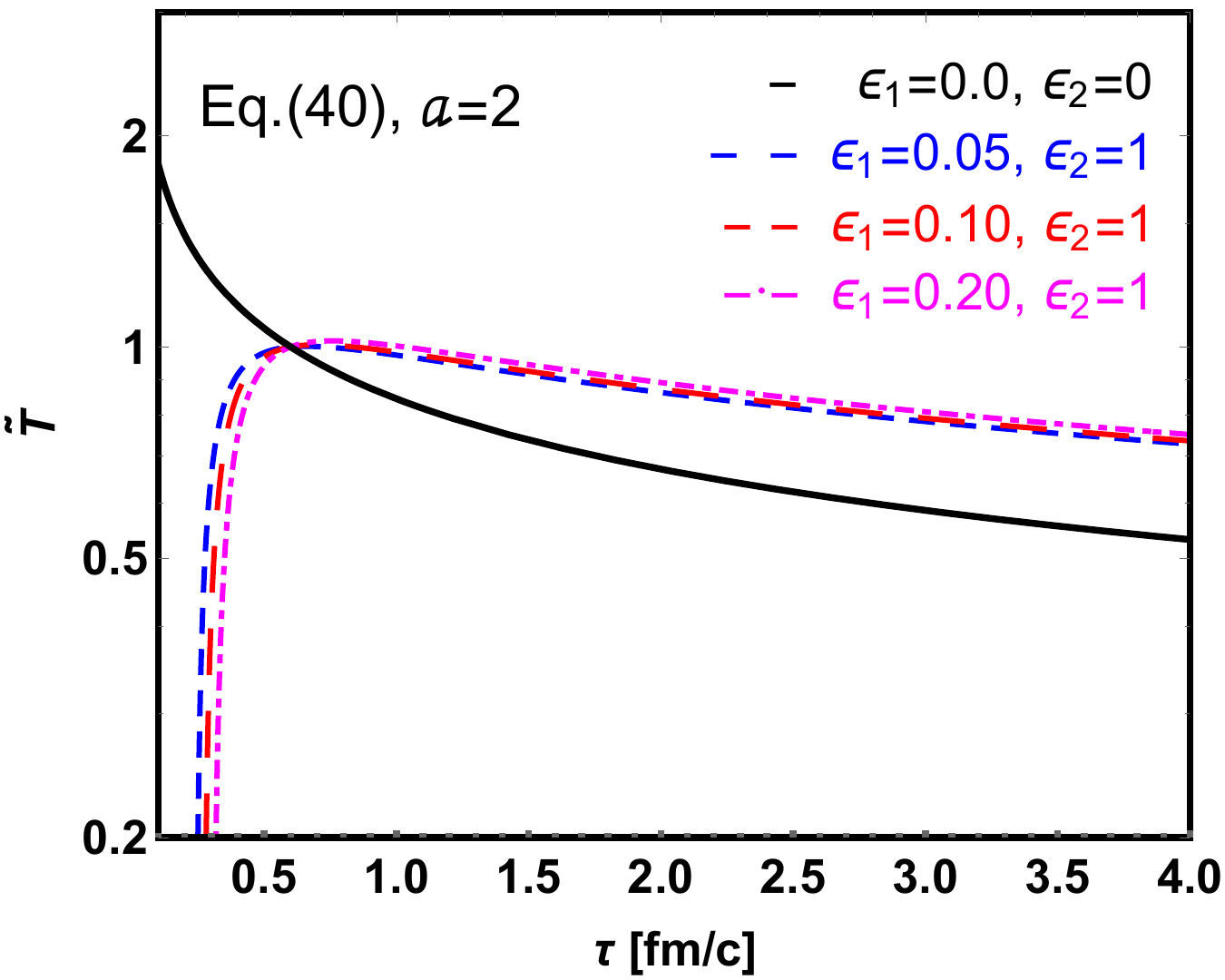}
\caption{(Color online) Evolution of normalized temperature $\widetilde{T}_{\textrm{NS-1}}$ (Eq.~(\ref{eq:40})) as a function of proper time $\tau$ for different initial magnetic field ($\epsilon_{1}$) and fixed shear viscosity value $\epsilon_{2}=1$, compared to the ideal MHD. The magnetic field decay parameter $a=2$. }
\label{f:ns-2}
\end{figure}
%%%%%%%%%%%%%%%%%%%%%%%%%% Figure %%%%%%%%%%%%%%%%%%%%%%%%

In this section, we present a well-developed solution for first-order Navier-Stokes viscous flow (\textit{Azwinndini-Bjorken} flow)~\cite{Muronga:2001zk}.
Starting with the energy conservation equation~(Eq.(\ref{energy_conservation_1})), and by setting the magnetic field contribution to zero ($\epsilon_{1}=0$), one obtains
\begin{equation}
\begin{aligned}
\frac{\partial \varepsilon}{\partial \tau}  + \frac{\varepsilon+p}{\tau} - \frac{1}{\tau}\pi - \Pi\frac{1}{\tau} &= 0, \\
%\Rightarrow \frac{\partial (3a_{1}T^{4})}{\partial \tau} + \frac{4a_{1}T^{4}}{\tau} - \frac{\pi}{\tau} - \frac{\Pi}{\tau} &= 0, \\
%\Rightarrow 12a_{1} T^{3}\frac{\partial T}{\partial \tau} + \frac{4a_{1}T^{4}}{\tau} - \frac{4\eta}{3\tau^{2}} -\frac{\zeta}{\tau^{2}} &=0,\\
\Rightarrow \frac{\partial T}{\partial \tau} + \frac{T}{3\tau} - \frac{\eta}{9 a_{1} T^{3} \tau^{2}} -\frac{\zeta}{12 a_{1}T^{3}\tau^{2}} &=0,\\
%\Rightarrow \frac{\partial T}{\partial \tau} + \frac{T}{3\tau} - \frac{b_{1}}{9 a_{1} \tau^{2}} -\frac{b_{2}}{12 a_{1} \tau^{2}} &=0, \\
%\Rightarrow \frac{\partial T}{\partial \tau} + \frac{T}{3\tau} - \left( \frac{b_{1}}{9 a_{1}}+\frac{b_{2}}{12 a_{1}} \right)\frac{1}{\tau^{2}} &=0, \\
\Rightarrow \frac{\partial \Tilde{T}}{\partial \tau}  + \frac{\tilde{T}}{3\tau} - \frac{1}{T_{0}}\left( \frac{b_{1}}{9 a_{1}}+\frac{b_{2}}{12 a_{1}} \right)\frac{1}{\tau^{2}} &= 0,
\label{vis_energy_ns_0}
\end{aligned}
\end{equation}
where $\pi=\frac{4}{3}\frac{\eta}{\tau}$, $\Pi=\zeta \frac{1}{\tau}$, $\eta=b_{1}T^{3}$ and $\zeta=b_{2}T^{3}$. For simplicity, we let $\epsilon_{2}=\frac{1}{T_{0}}\frac{b_{1}}{9 a_{1}}=\frac{4}{9T_{0}}\frac{\eta}{s}$ where bulk viscosity coefficient $\zeta=0$, then Eq.~(\ref{vis_energy_ns_0}) can be written as
\begin{equation}
\begin{aligned}
\frac{\partial \Tilde{T}}{\partial \tau}  + \frac{\tilde{T}}{3\tau} - \frac{\epsilon_{2}}{\tau^{2}} &= 0.
\label{vis_energy_ns_1}
\end{aligned}
\end{equation}
The solution of above Eq.~(\ref{vis_energy_ns_1}) is
\begin{equation}
\begin{aligned}
%\tilde{T} = \left(\frac{\tau_0}{\tau}\right)^{\frac{1}{3}} + \left(\frac{3\epsilon_{2}}{2\tau}\right) \left[ \left(\frac{\tau}{\tau_0}\right)^{\frac{2}{3}} - 1 \right], \\
\tilde{T} = \left(\frac{\tau_0}{\tau}\right)^{\frac{1}{3}} \left[1+ \frac{3\epsilon_{2}}{2\tau_{0}} \left( 1-\left(\frac{\tau_0}{\tau}\right)^\frac{2}{3} \right)\right].
%\Rightarrow\tilde{T} = \left(\frac{\tau_0}{\tau}\right)^{\frac{1}{3}} + \left(\frac{\left( \frac{b_{1}}{3 a_{1}}+\frac{b_{2}}{4 a_{1}} \right)}{2\tau T_{0}}\right) \left[ \left(\frac{\tau}{\tau_0}\right)^{\frac{2}{3}} - 1 \right] \\
%\Rightarrow\tilde{T} = \left(\frac{\tau_0}{\tau}\right)^{\frac{1}{3}} + \left(\frac{\left( \frac{4 \eta }{3 s }+\frac{\zeta}{s} \right)}{2\tau T_{0}}\right) \left[ \left(\frac{\tau}{\tau_0}\right)^{\frac{2}{3}} - 1 \right]
\end{aligned}
\end{equation}
This solution type is as same as the Azwinndini-Bjorken solution~\cite{Muronga:2003ta}. Note that a nonvanishing shear viscosity parameter $\epsilon_{2}$ makes the temperature cooling rate smaller.

In Fig.~\ref{f:ns-1}, we present the normalized dimensionless temperature $\Tilde{T}$ as a function of proper time $\tau$, considering various shear viscosity. We find with a positive shear viscosity, while $\epsilon_{2}$ increase from 0 to 2, $\Tilde{T}$ decreases slower than the ideal fluid case. Additionally, there is a peak in $\Tilde{T}$ in the case of NS approximation when $\epsilon_{2} \neq 0$.

\subsection{Perturbative analytical Solution for the NS approximation of viscous MHD - I}
\label{section-2-c}

In this section, we employ the nonconserved charges method~\cite{Siddique:2019gqh} to solve the energy conservation equation and derive a perturbative solution that accounts for both magnetic field ($\epsilon_{1}$) and shear viscosity ($\epsilon_{2}$).

The energy conservation equation for a fluid with non-zero shear viscosity and non-zero magnetic field with longitudinal boost invariance takes the following form
\begin{equation}
\begin{aligned}
\frac{\partial \left(\varepsilon + \frac{1}{2}B^{2}\right)}{\partial \tau}  + \frac{\varepsilon+p+B^{2}}{\tau} - \frac{1}{\tau}\pi - \Pi\frac{1}{\tau} &= 0, \\
\Rightarrow \frac{\partial \Tilde{T}}{\partial \tau}  + \frac{\tilde{T}}{3\tau} - \frac{\epsilon_1}{\tilde{T}^{3}}\left(\frac{\tau_{0}}{\tau}\right)^{2a}\frac{1}{\tau} - \frac{\epsilon_{2}}{\tau^{2}} &= 0.
\label{vis_energy_vmhd-2c}
\end{aligned}
\end{equation}
Here $\epsilon_{1}=\frac{(a-1)\sigma}{12 a_{1}}$ and $\epsilon_{2}=\frac{4}{T_{0}}\frac{b_{1}}{9 a_{1}}$.
One can not get the analytical solution of above equation directly. However, a perturbative solution is able to find through perturbative method.

In Ref.~\cite{Siddique:2019gqh}, a nonconserved charge method is used to solve the equation for $f\left(\tau\right)$ in the following form
\begin{eqnarray}
\frac{d}{d\tau}f\left(\tau\right)+m\frac{f\left(\tau\right)}{\tau}=f\left(\tau\right)\frac{d}{d\tau}\lambda\left(\tau\right),
\label{29}
\end{eqnarray}
where $m$ is a constant and $\lambda\left(\tau\right)$ is a known function. The general solution of above equation Eq.~(\ref{29}) is
\begin{eqnarray}
f\left(\tau\right)=f\left(\tau_{0}\right)\exp\left[\lambda\left(\tau\right)-\lambda\left(\tau_{0}\right)\right]\left(\frac{\tau_{0}}{\tau}\right)^{m},
\label{30}
\end{eqnarray}
where $\tau_0$ is an initial proper time and $f(\tau_{0})$ is determined by an initial value at $\tau_0$.

We first rewrite the energy conservation Eq.~\eqref{vis_energy_vmhd-2c} as follow
\begin{eqnarray}
\frac{d}{d\tau}\widetilde{T}+\frac{1}{3}\frac{\widetilde{T}}{\tau}=\widetilde{T}\frac{d}{d\tau}\lambda,
\label{4}
\end{eqnarray}
where $m=\frac{1}{3}$ and
\begin{eqnarray}
\frac{d}{d\tau}\lambda=\frac{\epsilon_{1}}{\widetilde{T}^{4}}\left(\frac{\tau_{0}}{\tau}\right)^{2a}\frac{1}{\tau}+\frac{1}{\widetilde{T}}\frac{\epsilon_{2}}{\tau^{2}}.
\label{5}
\end{eqnarray}
Following Eqs.~(\ref{29}-\ref{30}), the formal solution of Eq.~(\ref{4}) is
\begin{equation}
\begin{aligned}
\widetilde{T} &=\widetilde{T}\left(\tau_{0}\right)\exp\left[\lambda\left(\tau\right)-\lambda\left(\tau_{0}\right)\right]\left(\frac{\tau_{0}}{\tau}\right)^{\frac{1}{3}}=\left(\frac{\tau_{0}}{\tau}\right)^{\frac{1}{3}}x\left(\tau\right),
\label{6}
\end{aligned}
\end{equation}
with the condition $\widetilde{T}\left(\tau_{0}\right)=1$ and by introducing the variables
\begin{eqnarray}
x\left(\tau\right)=\exp\left[\lambda\left(\tau\right)-\lambda\left(\tau_{0}\right)\right].
\label{7}
\end{eqnarray}
With the variables $x\left(\tau\right)$, we have $dx(\tau)=xd\lambda(\tau)$ and we can rewrite the Eq.~\eqref{5} in the form of
\begin{eqnarray}
\frac{d}{d\tau}x=x\frac{\epsilon_{1}}{\widetilde{T}^{4}}\left(\frac{\tau_{0}}{\tau}\right)^{2a}\frac{1}{\tau}+x\frac{1}{\widetilde{T}}\frac{\epsilon_{2}}{\tau^{2}},
\label{8}
\end{eqnarray}
where $x\left(\tau_{0}\right)=1$.

We assume the magnetic field value $\epsilon_{1}$ as perturbation and solve them in powers of $\epsilon_{1}$. In the order of $\mathcal{O}\left(\epsilon_{1}^{0}\right)$, we have
\begin{eqnarray}
\frac{d}{d\tau}x=x\frac{1}{\widetilde{T}}\frac{\epsilon_{2}}{\tau^{2}}=\epsilon_{2}\left(\frac{1}{\tau_{0}}\right)^{\frac{1}{3}}\tau^{-\frac{5}{3}},
\label{9}
\end{eqnarray}
then the solution of zero-order can be written as
\begin{equation}
\begin{aligned}
x_{0}&=1+\int_{\tau_{0}}^{\tau}\epsilon_{2}\left(\frac{1}{\tau_{0}}\right)^{\frac{1}{3}}\tau_{1}^{-\frac{5}{3}}d\tau_{1} \\
&=1-\frac{3}{2}\epsilon_{2}\left(\frac{1}{\tau_{0}}\right)^{1/3}\tau_{1}^{-2/3}\bigg|_{\tau_{1}=\tau_{0}}^{\tau_{1}=\tau} \\
&=1+\frac{3\epsilon_{2}}{2\tau_{0}}\left[1-\left(\frac{\tau_{0}}{\tau}\right)^{2/3}\right].
\label{10}
\end{aligned}
\end{equation}
In the first order of $\mathcal{O}\left(\epsilon_{1}^{1}\right)$, we have
\begin{equation}
\begin{aligned}
\frac{d}{d\tau}x & = x\frac{\epsilon_{1}}{\widetilde{T}^{4}}\left(\frac{\tau_{0}}{\tau}\right)^{2a}\frac{1}{\tau}+x\frac{1}{\widetilde{T}}\frac{\epsilon_{2}}{\tau^{2}} \\
& =\frac{\epsilon_{1}}{x^{3}\tau}\left(\frac{\tau_{0}}{\tau}\right)^{2a-\frac{4}{3}}+\epsilon_{2}\left(\frac{1}{\tau_{0}}\right)^{\frac{1}{3}}\tau^{-\frac{5}{3}},
\label{11}
\end{aligned}
\end{equation}
then we obtain
\begin{widetext}
\begin{equation}
\centering
\begin{footnotesize}
\begin{aligned}x_{1}= & x_{0}+\int_{\tau_{0}}^{\tau}\frac{\epsilon_{1}}{x\left(\tau_{1}\right)^{3}\tau_{1}}\left(\frac{\tau_{0}}{\tau_{1}}\right)^{2a-\frac{4}{3}}d\tau_{1}\\
= & x_{0}+\int_{\tau_{0}}^{\tau}\frac{\epsilon_{1}}{\left[1+\frac{3\epsilon_{2}}{2\tau_{0}}\left(1-\left(\frac{\tau_{0}}{\tau_{1}}\right)^{2/3}\right)\right]^{3}\tau_{1}}\left(\frac{\tau_{0}}{\tau_{1}}\right)^{2a-\frac{4}{3}}d\tau_{1}\\
= & 1+\frac{3\epsilon_{2}}{2\tau_{0}}\left[1-\left(\frac{\tau_{0}}{\tau}\right)^{2/3}\right]+\epsilon_{1}\left\{ \int_{\tau_{0}}^{\infty}\frac{1}{\left[1+\frac{3\epsilon_{2}}{2\tau_{0}}\left(1-\text{\ensuremath{\Bigl(\frac{\tau_{0}}{\tau_{1}}\Bigr)}}^{2/3}\right)\right]^{3}\tau_{1}}\Bigl(\frac{\tau_{0}}{\tau_{1}}\Bigr)^{2a-\frac{4}{3}}d\tau_{1}-\int_{\tau}^{\infty}\frac{1}{\left[1+\frac{3\epsilon_{2}}{2\tau_{0}}\left(1-\Bigl(\frac{\tau_{0}}{\tau_{1}}\Bigr)^{2/3}\right)\right]^{3}\tau_{1}}\Bigl(\frac{\tau_{0}}{\tau_{1}}\Bigr)^{2a-\frac{4}{3}}d\tau_{1}\right\} \\
= & 1+\frac{3\epsilon_{2}}{2\tau_{0}}\left[1-\left(\frac{\tau_{0}}{\tau}\right)^{2/3}\right]+\epsilon_{1}\left\{ \int_{1}^{\infty}\frac{1}{\left[1+\frac{3\epsilon_{2}}{2\tau_{0}}\left(1-\Bigl(\frac{1}{\tau_{2}}\Bigr)^{2/3}\right)\right]^{3}\tau_{2}}\Bigl(\frac{1}{\tau_{2}}\Bigr)^{2a-\frac{4}{3}}d\tau_{2}-\int_{1}^{\infty}\frac{1}{\left[1+\frac{3\epsilon_{2}}{2\tau_{0}}\left(1-\Bigl(\frac{\tau_{0}}{\tau_{2}\tau}\Bigr)^{2/3}\right)\right]^{3}\tau_{2}}\Bigl(\frac{\tau_{0}}{\tau_{2}\tau}\Bigr)^{2a-\frac{4}{3}}d\tau_{2}\right\} \\
= & 1+\frac{3\epsilon_{2}}{2\tau_{0}}\left[1-\left(\frac{\tau_{0}}{\tau}\right)^{2/3}\right]+\frac{3\tau_{0}\epsilon_{1}}{2\left(2\tau_{0}+3\epsilon_{2}\right)^{3}}\Biggl\{\frac{4\tau\tau_{0}}{\left(2-3a\right)\Bigl[2\tau_{0}-3\epsilon_{2}\Bigl(\left(\frac{\tau_{0}}{\tau}\right)^{2/3}-1\Bigr)\Bigr]^{2}}\Bigl(\frac{\tau_{0}}{\tau}\Bigr)^{2a-\frac{1}{3}}\biggl[-9a^{2}\left(2\tau_{0}+3\epsilon_{2}\right)^{2}+3\left(a-1\right)\left(3a-4\right)\\
 & \times\Bigl[2\tau_{0}-3\epsilon_{2}\Bigl(\left(\frac{\tau0}{\tau}\right)^{2/3}-1\Bigr)\Bigr]^{2}\,_{2}F_{1}\left(1,3a-2;3a-1;\frac{3\epsilon_{2}}{3\epsilon_{2}+2\tau_{0}}\left(\frac{\tau_{0}}{\tau}\right)^{2/3}\right)+6\left(4-3a\right)\epsilon_{2}\left(\frac{\tau_{0}}{\tau}\right)^{2/3}\left(2\tau_{0}+3\epsilon_{2}\right)+9a\left(3a-4\right)\epsilon_{2}\left(\frac{\tau_{0}}{\tau}\right)^{2/3}\\
 & \times\left(2\tau_{0}+3\epsilon_{2}\right)+21a\left(2\tau_{0}+3\epsilon_{2}\right)^{2}-10\left(2\tau_{0}+3\epsilon_{2}\right)^{2}\biggr]+12\left(a-1\right)\left(3a-4\right)\tau_{0}^{2}\Gamma\left(3a-2\right)\,_{2}\widetilde{F}_{1}\left(1,3a-2;3a-1;\frac{3\epsilon_{2}}{3\epsilon_{2}+2\tau_{0}}\right)\\
 & +\left(2\tau_{0}+3\epsilon_{2}\right)\Bigl[2\left(5-3a\right)\tau_{0}+3a\epsilon_{2}\Bigr]\Biggr\},
\end{aligned}
\label{12}
\end{footnotesize}
\end{equation}
\end{widetext}
where $\,_{2}F_{1}\left(a,b;c;z\right)$ is hypergeometric function, $\Gamma(z)$ is Gamma function, and $\,_{2}\widetilde{F}_{1}\left(a,b;c;z\right)$ is regularized hypergeometric function.

Consequently, the perturbative solution of Eq.~\eqref{vis_energy_vmhd-2c} is given by
\begin{eqnarray}
\widetilde{T}_{\textrm{NS-1}}=\left(\frac{\tau_{0}}{\tau}\right)^{1/3}x_{1},
\label{eq:40}
\end{eqnarray}
where $x_{1}$ is the Eq.~(\ref{12}), and this solution is stable where $\epsilon_1$ is small. We will prove this in the next section~\ref{sec-4-1}.

In Fig.~\ref{f:ns-2}, we present the normalized temperature $\widetilde{T}_{\textrm{NS-1}}$ (given by Eq.~(\ref{eq:40})) as a function of proper time, for different initial magnetic field strengths ($\epsilon_{1}$) and a fixed shear viscosity ($\epsilon_{2}=1$). The magnetic field decay parameter is set to $a=2$, and the initial proper time is $\tau_{0}=0.6$ fm/c.
One finds that as $\epsilon_{1}$ increases from 0.05 to 0.2, the temperature decreases more slowly compared to the ideal MHD case. Furthermore, for non-zero magnetic field, the fluid will absorb energy in excess of the decay caused by the expansion, and leads to a peak for proper times $\tau<0.6$ fm/c~\cite{Roy:2015kma}.

\subsection{Perturbative analytical solution for the NS approximation of viscous MHD - II}
\label{section-2-d}

%%%%%%%%%%%%%%%%%%% Fig -4 %%%%%%%%%%%%%%%%%%%%%%%%%%
\begin{figure}[tbp!]
\includegraphics[width=0.85\linewidth]{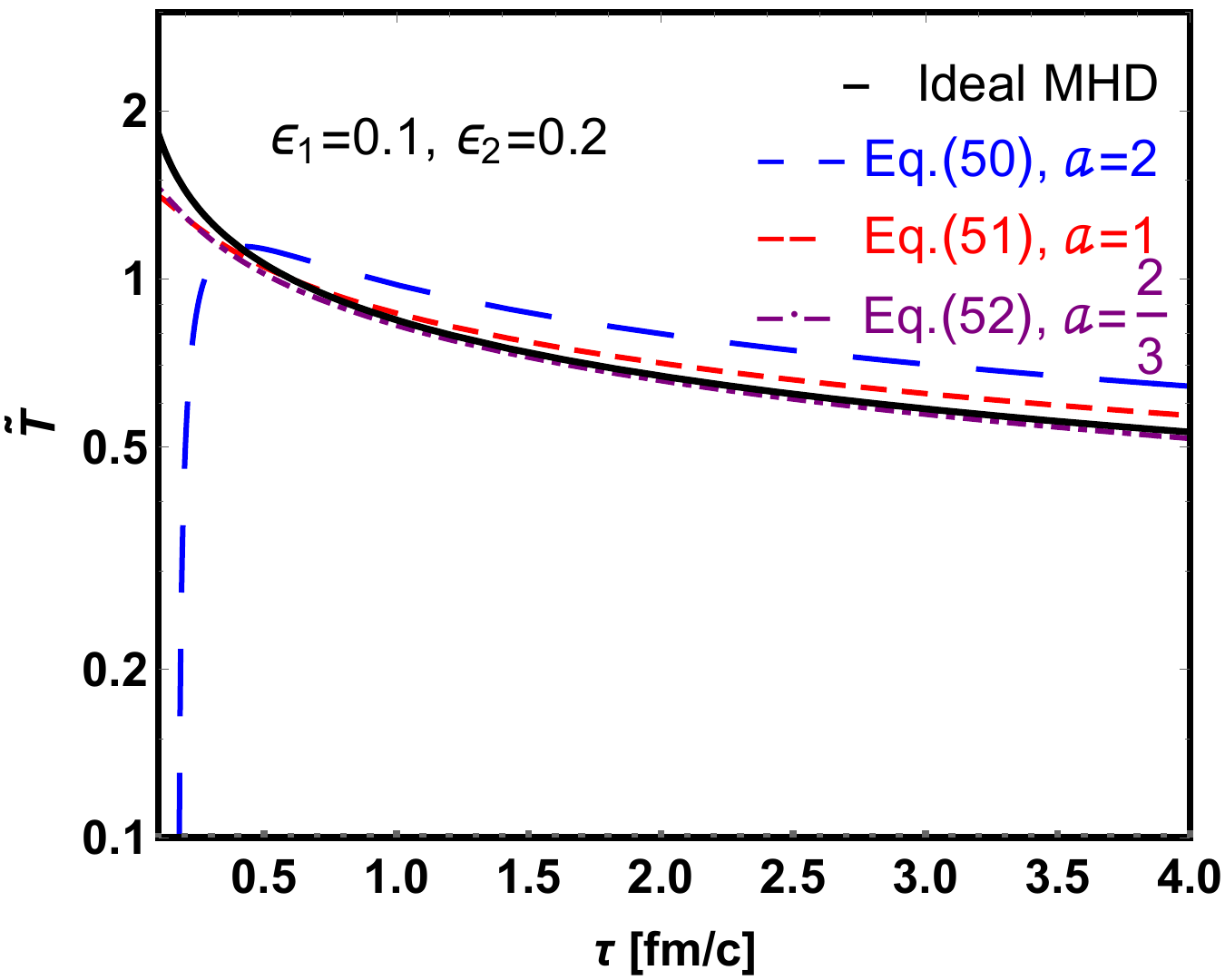}
\caption{(Color online) Evolution of normalized temperature $\widetilde{T}_{\textrm{NS-2}}$ (Eqs.~(\ref{eq:50}-\ref{eq:52})) as a function of proper time $\tau$ for different values of magnetic decay parameter $a$ with fixed initial magnetic field ($\epsilon_{1}=0.1$) and fixed shear viscosity value ($\epsilon_{2}=0.2$), compared to the ideal MHD. }
\label{f:ns-3}
\end{figure}
%%%%%%%%%%%%%%%%%%%%%%%%%%%%%%%%%%%%%%%%%%%%%

In the previous subsection-\ref{section-2-c}, we introduced a perturbative method that can be utilized to derive an analytical solution for the NS approximation of viscous fluid dynamical equations, under the assumption that the magnetic field is small ($\epsilon_{1}$ being a small parameter).
In this section, we present another perturbative solution for viscous non-resistive MHD flow with NS approximation, assuming that both $\epsilon_{1}$ and $\epsilon_{2}$ are small parameters.

The energy conservation equation (Eq.~(\ref{vis_energy_vmhd-2c})) for viscous MHD flow under NS approximation is
\begin{equation}
\begin{aligned}
\frac{\partial \Tilde{T}}{\partial \tau}  + \frac{\tilde{T}}{3\tau} - \frac{\epsilon_{2}}{\tau^{2}} &= \epsilon_1\frac{1}{\tilde{T}^{3}}\left(\frac{\tau_{0}}{\tau}\right)^{2a}\frac{1}{\tau}.
\label{vis_energy}
\end{aligned}
\end{equation}
Here we assume that $\epsilon_{1}$ is a small parameter. To the linear order in $\epsilon_{1}$, the temperature $\Tilde{T}$ can be written as
\begin{equation}
\begin{aligned}
\tilde{T} = \tilde{T}_{0} + \epsilon_{1} \tilde{T}_{1} + \epsilon_{1}^{2} \tilde{T}_{2} + ...,
\label{pert}
\end{aligned}
\end{equation}
putting Eq.~(\ref{pert}) into Eq.~(\ref{vis_energy}), one obtains

\begin{equation}
\begin{aligned}
&\frac{\partial (\tilde{T}_{0} + \epsilon_{1} \tilde{T}_{1}  + \epsilon_{1}^{2} \tilde{T}_{2} +... )}{\partial \tau}  + \frac{\tilde{T}_{0} + \epsilon_{1} \tilde{T}_{1}  + \epsilon_{1}^{2} \tilde{T}_{2} +... }{3\tau} - \frac{\epsilon_{2}}{\tau^{2}} \\
&~~~~~~~~~~~~~~= \frac{\epsilon_1}{(\tilde{T}_{0} + \epsilon_{1} \tilde{T}_{1}  + \epsilon_{1}^{2} \tilde{T}_{2} +...)^{3}}\left(\frac{\tau_{0}}{\tau}\right)^{2a}\frac{1}{\tau}.
\label{vis_energy_tot}
\end{aligned}
\end{equation}

\begin{figure*}[tbp!]
\includegraphics[width=0.4\linewidth]{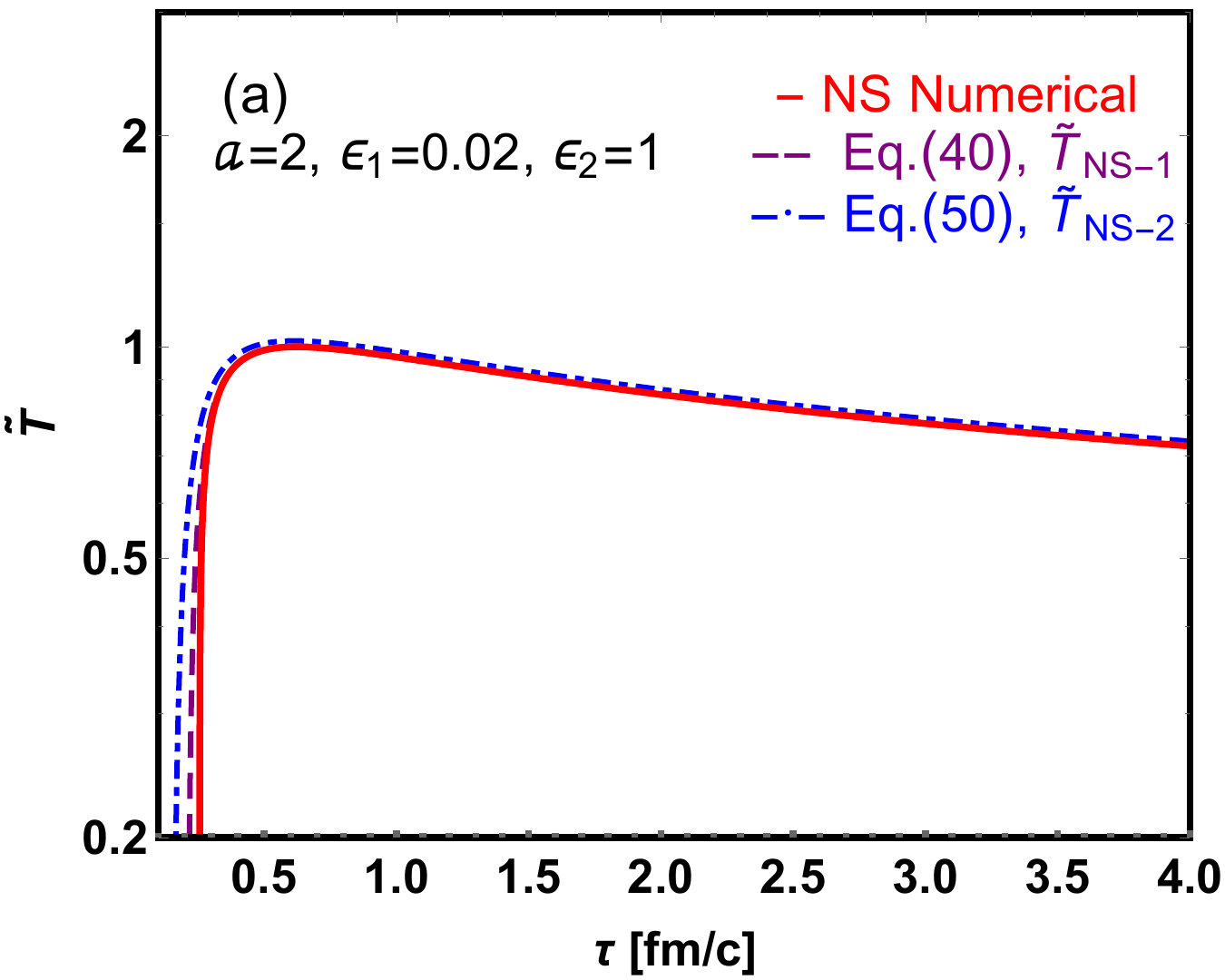}~~
\includegraphics[width=0.4\linewidth]{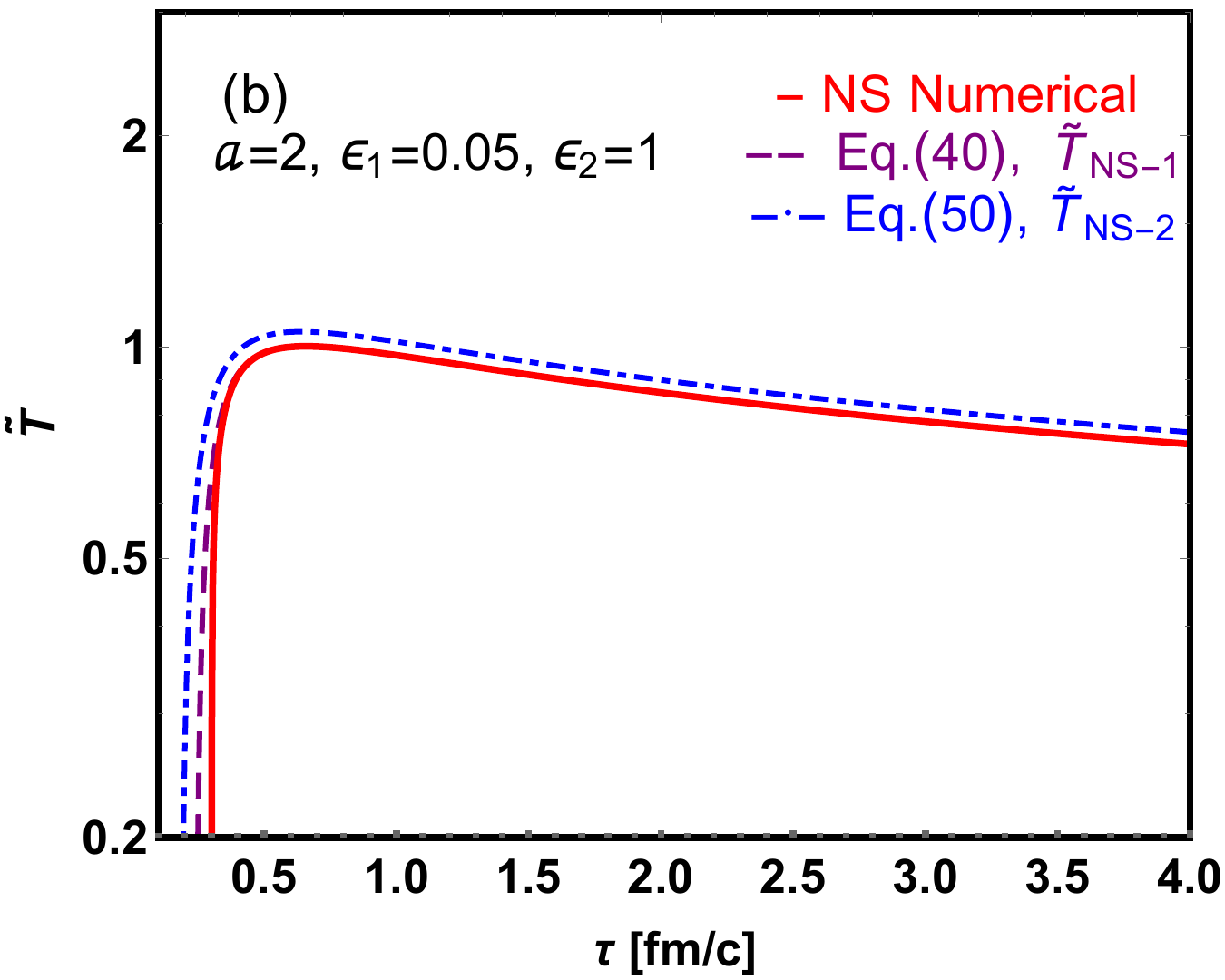} \\
\includegraphics[width=0.4\linewidth]{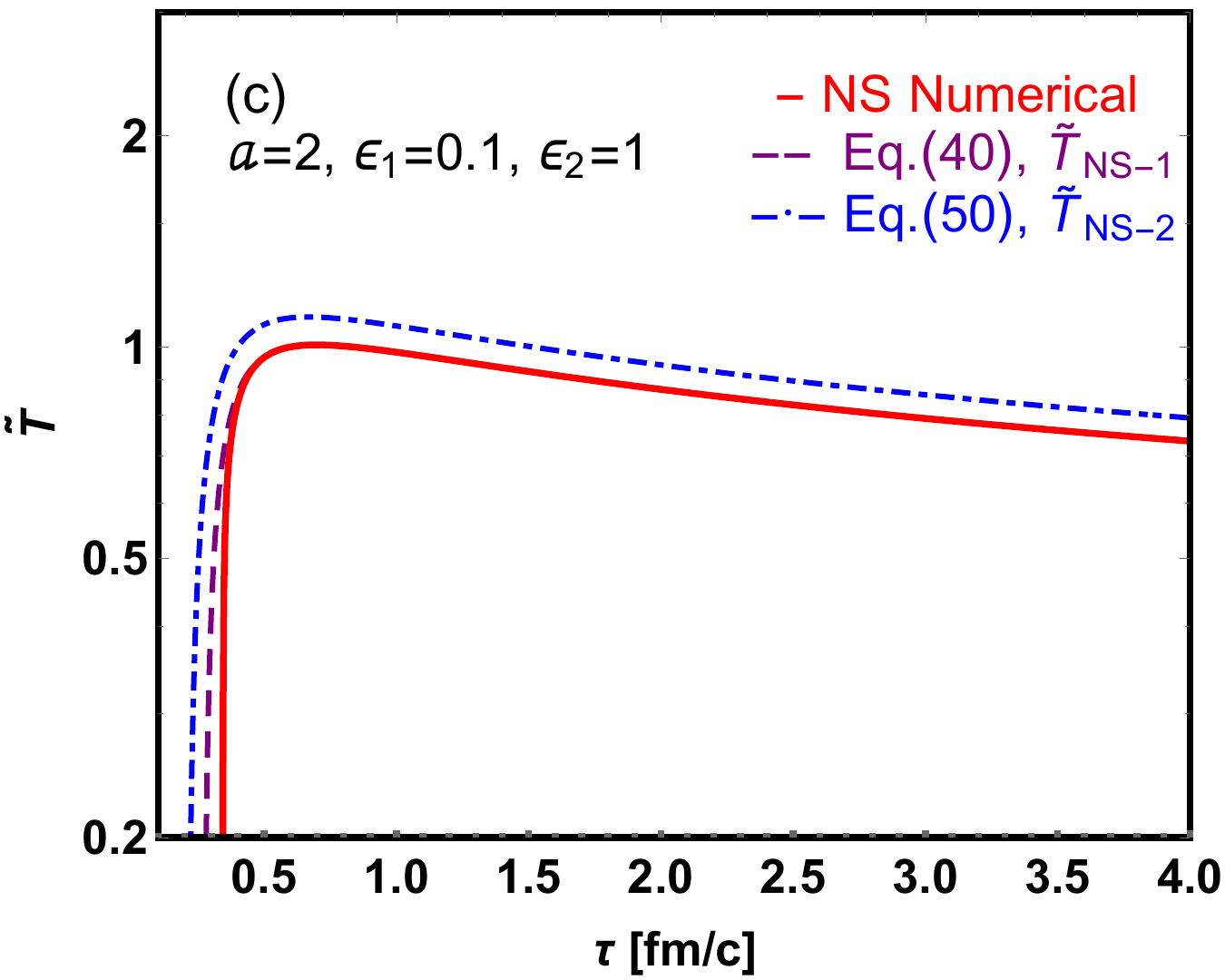}~~
\includegraphics[width=0.4\linewidth]{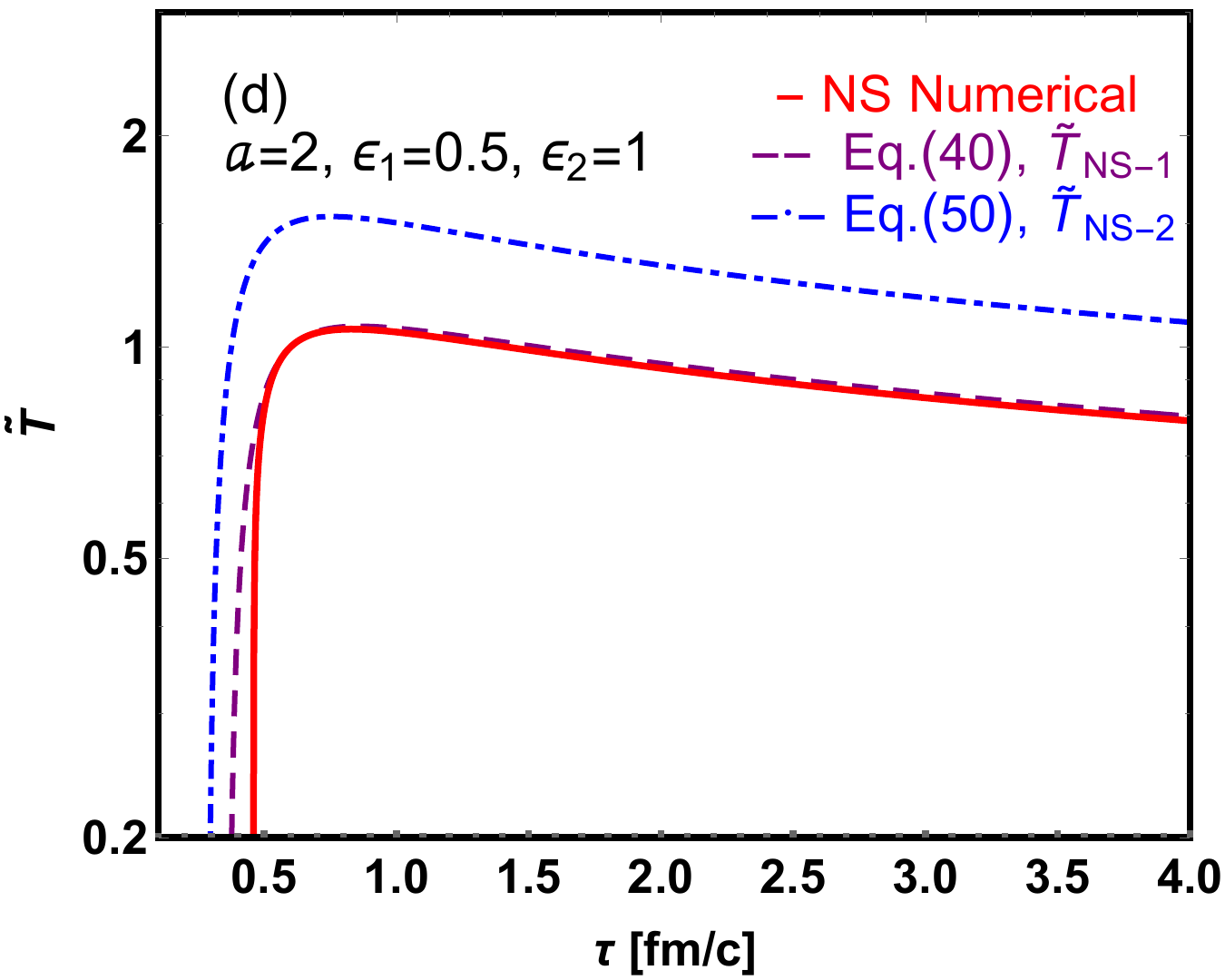} \\
\caption{(Color online) Evolution of normalized temperature $\tilde{T}$ as a function of proper time $\tau$, comparing two perturbative analytical solutions with numerical solutions. The normalized temperatures $\tilde{T}$ are presented for various values of the magnetic field parameter $\epsilon_{1}$, with a fixed shear viscosity ($\epsilon_{2}=1$) and a magnetic field decay parameter $a=2$. }
\label{f:numerical-1st-1}
\end{figure*}

For simplicity, we ignore terms higher than $\mathcal{O}(\epsilon_{1}^{2})$. Consequently, Eq.~(\ref{vis_energy_tot}) reduces to
\begin{equation}
\begin{aligned}
\frac{\partial (\tilde{T}_{0} + \epsilon_{1} \tilde{T}_{1})}{\partial \tau}  &+ \frac{\tilde{T}_{0} + \epsilon_{1} \tilde{T}_{1}}{3\tau} - \frac{\epsilon_{2}}{\tau^{2}} \\
&~~~~ =\frac{\epsilon_1}{(\tilde{T}_{0}^{3}+3\epsilon_{1} \tilde{T}_{0}^{2}\tilde{T}_{1})}\left(\frac{\tau_{0}}{\tau}\right)^{2a}\frac{1}{\tau}.
\label{vis_1term}
\end{aligned}
\end{equation}
By combining the like terms in the above Eq.~(\ref{vis_1term}), we obtain
\begin{equation}
\left\{
\begin{aligned}
%\nonumber
\frac{\partial \tilde{T}_{0}}{\partial \tau}  + \frac{\tilde{T}_{0}}{3\tau} - \frac{\epsilon_{2}}{\tau^{2}} &= 0, \\
\frac{\partial (\epsilon_{1} \tilde{T}_{1})}{\partial \tau}  + \frac{\epsilon_{1} \tilde{T}_{1}}{3\tau}  &= \frac{\epsilon_1}{(\tilde{T}_{0}^{3}+3\epsilon_{1} \tilde{T}_{0}^{2}\tilde{T}_{1})}\left(\frac{\tau_{0}}{\tau}\right)^{2a}\frac{1}{\tau}.
\end{aligned}
\right.
\label{vis_energy_12}
\end{equation}
The analytical solution to the first equation in Eq.~(\ref{vis_energy_12}) is
\begin{equation}
\begin{aligned}
\tilde{T}_{0} = \left(\frac{\tau_0}{\tau}\right)^{\frac{1}{3}} + \left(\frac{3\epsilon_{2}}{2\tau}\right) \left[ \left(\frac{\tau}{\tau_0}\right)^{\frac{2}{3}} - 1 \right].
\label{solu_t0}
\end{aligned}
\end{equation}
Putting $\tilde{T_{0}}$ into the second equation of Eq.~(\ref{vis_energy_12}), we arrive at
\begin{equation}
\begin{aligned}
\frac{\partial \tilde{T}_{1}}{\partial \tau}  + \frac{\tilde{T}_{1}}{3\tau} &= \frac{1}{(\tilde{T}_{0}^{3}+3\epsilon_{1} \tilde{T}_{0}^{2}\tilde{T}_{1})}\left(\frac{\tau_{0}}{\tau}\right)^{2a}\frac{1}{\tau}.
\label{vis_energy_13}
\end{aligned}
\end{equation}
To obtain an analytical solution of Eq.~(\ref{vis_energy_13}), we approximate $\tilde{T}_{0}^{3}(1+3\epsilon_{1} \tilde{T}_{0}/\tilde{T}_{1})\approx \tilde{T_{0}}^{3}\approx(\tau_{0}/\tau$), where $\epsilon_{1}$ and $\epsilon_{2}$ are small parameter. We note that such an approximation has very significant limitations, and we will present its limitation later in Fig.~\ref{f:numerical-1st-1}.
Accordingly, Eq.~(\ref{vis_energy_13}) reduces to
\begin{equation}
\begin{aligned}
\frac{\partial \tilde{T}_{1}}{\partial \tau}  + \frac{\tilde{T_{1}}}{3\tau} &= \left(\frac{\tau_{0}}{\tau}\right)^{2a-1}\frac{1}{\tau}.
\label{vis_energy_14}
\end{aligned}
\end{equation}
We obtain an analytical solution of Eq.~(\ref{vis_energy_14}) as follow
\begin{equation}
\begin{aligned}
\tilde{T_{1}} = \frac{1}{4-6a} \left(\frac{\tau_{0}}{\tau}\right)^{\frac{1}{3}} \left[(1-6a) + 3\left(\frac{\tau_{0}}{\tau}\right)^{2a-\frac{4}{3}}\right].
\end{aligned}
\end{equation}

Combining the zero-order ($\tilde{T_{0}}$) and first order terms ($\tilde{T_{1}}$), we obtain a perturbative analytical solution as follow,
\begin{equation}
\begin{aligned}
\tilde{T}_{\textrm{NS-2}} &= \tilde{T}_{0} + \epsilon_{1}\tilde{T}_{1} \\
%&= \left(\frac{\tau_0}{\tau}\right)^{\frac{1}{3}} +  \left(\frac{3\epsilon_{2}}{2\tau}\right) \left[ \left(\frac{\tau}{\tau_0}\right)^{\frac{2}{3}} - 1 \right] \\
%&~~~~+ \frac{\epsilon_{1}}{4-6a} \left(\frac{\tau_{0}}{\tau}\right)^{\frac{1}{3}} \left[(1-6a) + 3\left(\frac{\tau_{0}}{\tau}\right)^{2a-\frac{4}{3}}\right]\\
&= \left(\frac{\tau_0}{\tau}\right)^{\frac{1}{3}} \Big{[} 1+ \frac{3\epsilon_{2}}{2\tau_{0}} \left( 1-\left(\frac{\tau_0}{\tau}\right)^\frac{2}{3} \right) \\
&~~~~+ \frac{\epsilon_{1}}{4-6a}\left( 1-6a+3\left(\frac{\tau_{0}}{\tau}\right)^{2a-\frac{4}{3}} \right) \Big{]}.
\label{eq:50}
\end{aligned}
\end{equation}
Above solution~Eq.(\ref{eq:50}) exhibits divergent behavior for different values of the magnetic field decay parameter $a$. We demonstrate following three particular limits as we studied in previous subsection-\ref{section-2-a}.

\emph{\textbf{Case-A}} For $a=1$, we obtain a solution the same as the NS approximation for viscous flow.
\begin{equation}
\begin{aligned}
\tilde{T}_{\textrm{NS-2}} &= \left(\frac{\tau_0}{\tau}\right)^{\frac{1}{3}} \left[ 1+ \frac{3\epsilon_{2}}{2\tau_{0}} \left( 1-\left(\frac{\tau_0}{\tau}\right)^\frac{2}{3} \right)\right].
\label{eq:51}
\end{aligned}
\end{equation}

\emph{\textbf{Case-B}} For $a \rightarrow \frac{2}{3}$ and $\tau>\tau_{0}$, we obtain a solution as follow
\begin{equation}
\begin{aligned}
\tilde{T}_{\textrm{NS-2}} &=\left( \frac{\tau_0}{\tau} \right)^{\frac{1}{3}} \left[ 1 + \frac{3 \epsilon_2}{2 \tau_0} \left( 1 - \left( \frac{\tau_0}{\tau} \right)^{\frac{2}{3}} \right) + \frac{\epsilon_{1}\left( \ln\left(\frac{\tau_0}{\tau}\right) - 1 \right)}{3(a-1)}  \right].
\label{eq:52}
\end{aligned}
\end{equation}

%%%%%%%%%%%%%%%%%%%%%%%%%%%%%%%%%%%%%%%%%%%%%%%%%%%%%%
%%%%%%%%%%%%%%%%%%%%%%%%%%%%%%%%%%%%%%%%%%%%%%%%%%%%%%%

\emph{\textbf{Case-C}} For $a \rightarrow \infty$ and $\tau>\tau_{0}$, we assume that initial magnetic strength parameter $\sigma$ is a first order infinitesimal quantity, specifically $\mathcal{O}(a)$, associated with $a$, one finds that the approximate solution regress to the viscous solution. Alternatively, if we consider $\sigma$ as a second-order small quantity, its resulting magnetic field effects become too negligible to achieve our desired effect.

In Fig.~\ref{f:ns-3}, we present the normalized temperature as a function of proper time for various magnetic decay rates ($a$), with a fixed initial value of the magnetic field ($\epsilon_{1}=0.1$) and shear viscosity ($\epsilon_{2}=0.2$). One finds for $0<a<1$, the third term within the $\Tilde{T}$ in Eq.~(\ref{eq:52}) is always negative. This results in a faster decay compared to both the ideal MHD case and the $a=1$ case (as described in Eq.~(\ref{eq:51})). Furthermore, when $a>1$, the third term of $\Tilde{T}$ in Eq.~(\ref{eq:50}) is always positive, results in a slower decay compared to the ideal MHD case.

\section{Numerical solution of viscous MHD flow}
\label{section-4}

In this section, we present the numerical solution of viscous non-resistive magnetohydrodynamics with longitudinal boost invariance with first order Navier-Stokes (NS) approximation and second order Muller-Israel-Stewart (IS) approximation theory.

\subsection{First order Navier-Stokes approximation}
\label{sec-4-1}
We begin with the energy conservation equation with first order NS approximation as follow,
\begin{equation}
\begin{aligned}
\frac{\partial \Tilde{T}}{\partial \tau}  + \frac{\tilde{T}}{3\tau} - \frac{\epsilon_1}{\tilde{T}^{3}}\left(\frac{\tau_{0}}{\tau}\right)^{2a}\frac{1}{\tau} - \frac{\epsilon_{2}}{\tau^{2}} &= 0.
\label{eq:53}
\end{aligned}
\end{equation}
With the initial condition $T_{0}=0.65$ GeV and $\tau_{0}=0.6$ fm/c, we obtain the numerical solution of above Eq.~(\ref{eq:53}) quickly.

In Fig.~\ref{f:numerical-1st-1} we show the normalized temperature $\tilde{T}$ as a function of proper time $\tau$ with different magnetic field $\epsilon_{1}$ and fixed shear viscosity $\epsilon_{2}$, and compared it with the previous perturbative analytical solutions Eq.~(\ref{eq:40}) and Eq.~(\ref{eq:50}). The initial proper time sets $\tau_{0}=0.6$ fm/c, and the magnetic field decay parameter $a=2$.

In the upper left panel of Figure~\ref{f:numerical-1st-1}, with parameters $a=2$, $\epsilon_{1}=0.02$ and $\epsilon_{2}=1$, the initial strength of magnetic field is given by $B_{0}^{2}=0.24a_{1} T_{0}^{4}$. We observe that the analytical solution $\tilde{T}_{\textrm{NS-1}}$ aligns very closely with the numerical results, whereas the solution $\tilde{T}_{\textrm{NS-2}}$ is found to be in close to and slightly higher than the numerical solution.

In the upper right panel of Figure~\ref{f:numerical-1st-1}, for the same value for $a=2$ but with $\epsilon_{1}=0.05$ and $\epsilon_{2}=1$, the initial strength of magnetic field is calculated as $B_{0}^{2}=0.6a_{1} T_{0}^{4}$. Here, the solution $\tilde{T}_{\textrm{NS-1}}$ is again in good agreement with the numerical results, whereas the solution $\tilde{T}_{\textrm{NS-2}}$ deviates to become 5\% greater than the numerical solution.

In the lower panel of Fig.~\ref{f:numerical-1st-1}, with the magnetic field parameter $\epsilon_{1}=0.1$ in the left and $\epsilon_{1}=0.5$ in the right panel. We find that the analytical solution $\tilde{T}_{\textrm{NS-1}}$ remains consistent with the numerical solution. However, the analytical solution $\tilde{T}_{\textrm{NS-2}}$ deviates more significantly from the numerical solution as the $\epsilon_{1}$ increases, indicating that this analytical solution becomes less stable compared to the analytical solution $\tilde{T}_{\textrm{NS-1}}$ and numerical solution.

We summarize the differences between the two perturbative solutions (Eqs.~(\ref{eq:40}) and (\ref{eq:50})) as follows: (1) When the magnetic field parameter $\epsilon_{1}$ is smaller than 0.05, both perturbative analytical solutions exhibit the same distribution and reflect the main features of the temperature profile as derived from the numerical solution; (2) the solution $\tilde{T}_{\textrm{NS-1}}$ (Eq.~(\ref{eq:40})) performs well when the magnetic field parameter $\epsilon_{1}$ is small, and this is not significantly affected by the shear viscosity ($\epsilon_{2}$) of the fluid.

\subsection{Second Order Israel-Stewart approximation}

In the second order theory, the bulk viscous pressure $\Pi$ and shear viscosity $\pi$ have to be determined from the second order transport equations. Based on Refs.~\cite{Muronga:2001zk,Muronga:2003ta}, the energy conservation equation and viscosity transport equations can be written as
\begin{equation}
\begin{aligned}
\frac{\partial \left(\varepsilon + \frac{1}{2}B^{2}\right)}{\partial \tau}  + \frac{\varepsilon+p+B^{2}}{\tau} - \frac{1}{\tau}\pi - \Pi\frac{1}{\tau} = 0,
\label{eq:54}
\end{aligned}
\end{equation}

\begin{equation}
\begin{aligned}
\frac{\partial\Pi}{\partial\tau} = -\frac{\Pi}{\tau_{\Pi}} - \frac{1}{2}\frac{1}{\beta_{0}}\Pi\left[\beta_{0}\frac{1}{\tau} + T\frac{\partial}{\partial\tau} \left(\frac{\beta_{0}}{T}\right)\right] - \frac{1}{\beta_{0}}\frac{1}{\tau},
\label{eq:55}
\end{aligned}
\end{equation}
\begin{equation}
\begin{aligned}
\frac{\partial\pi}{\partial\tau} = -\frac{\pi}{\tau_{\pi}} - \frac{1}{2}\pi\left[\frac{1}{\tau} + \frac{1}{\beta_{2}}T\frac{d}{d\tau} \left(\frac{\beta_{2}}{T}\right)\right] + \frac{2}{3}\frac{1}{\beta_{2}}\frac{1}{\tau}.
\label{eq:56}
\end{aligned}
\end{equation}

For massless particles, the relaxation coefficient $\beta_{2}=3/(4p)$. The relaxation time $\tau_{\pi}$ is then calculated as $\tau_{\pi}=2\eta\beta_{2}$~\cite{Muronga:2003ta}.
As we discussed in the previous sections, since we are considering a system of massless particles and utilizing the equation of state  $\varepsilon=3p$, this leads to the absence of bulk viscosity, and therefore, it will not be included in subsequent discussions.

Based on Ref.~\cite{Muronga:2003ta}, the value of initial shear stress (viscosity) $\pi_{0}$ ( defined as $\pi_{0}=\pi^{00}_{0}-\pi_{0}^{33}$) can be determined by the initial pressure $p_{0}$. For simplicity, we assume an initial condition where $\pi_{0} \equiv b_{3}p_{0}=a_{1}b_{3}T_{0}^{4}$, with $b_{3}$ being a constant. We further define a normalized fluid shear viscosity $\tilde{\pi}$ as $\tilde{\pi}=\pi/\pi_{0}$ and make it a dimensionless parameter. Consequently, the energy conservation equation simplifies to
\begin{equation}
\begin{aligned}
\frac{\partial \left(\varepsilon + \frac{1}{2}B^{2}\right)}{\partial \tau} &= -\frac{\varepsilon+p+B^{2}}{\tau} + \frac{1}{\tau}\pi, \\
%\Rightarrow \frac{\partial \Tilde{T}}{\partial \tau} &=  - \frac{\tilde{T}}{3\tau} + \frac{\epsilon_1}{\tilde{T}^{3}}\left(\frac{\tau_{0}}{\tau}\right)^{2a}\frac{1}{\tau} + \frac{\pi}{12a_{1} T^{3} T_{0} \tau}, \\
\Rightarrow\frac{\partial \Tilde{T}}{\partial \tau} &= - \frac{\tilde{T}}{3\tau} + \frac{\epsilon_1}{\tilde{T}^{3}}\left(\frac{\tau_{0}}{\tau}\right)^{2a}\frac{1}{\tau} + \frac{b_{3}\tilde{\pi}\tilde{T}}{12 \tau}.
\label{eq:57}
\end{aligned}
\end{equation}
And the shear viscosity equation reduces to
\begin{equation}
\begin{aligned}
\frac{\partial\pi}{\partial\tau} &= -\frac{\pi}{\tau_{\pi}} - \frac{1}{2}\pi\left[\frac{1}{\tau} + \frac{1}{\beta_{2}}T\frac{\partial}{\partial\tau} \left(\frac{\beta_{2}}{T}\right)\right] + \frac{2}{3}\frac{1}{\beta_{2}}\frac{1}{\tau}, \\
%\Rightarrow\frac{\partial\pi}{\partial\tau} &= -\frac{2\pi p}{3\eta} - \frac{\pi}{2}\left[\frac{1}{\tau} + \frac{4p}{3}T\frac{\partial}{\partial\tau} \left(\frac{3}{4p T}\right)\right] + \frac{2}{3}\frac{4p}{3}\frac{1}{\tau} \\
\Rightarrow\frac{\partial\pi}{\partial\tau} &= -\frac{2\pi a_{1}T^{4}}{3b_{1}T^{3}} - \frac{\pi}{2}\left[\frac{1}{\tau} + T^{5}\frac{\partial}{\partial\tau} \left(\frac{1}{T^{5}}\right)\right] + \frac{8a_{1}T^{4}}{9}\frac{1}{\tau}, \\
%\Rightarrow\frac{\partial\pi}{\partial\tau} &= -\frac{2a_{1}T \pi }{3b_{1}} - \frac{\pi}{2}\left[\frac{1}{\tau} - 5\frac{1}{T} \frac{\partial T}{\partial\tau}\right] + \frac{8a_{1}T^{4}}{9\tau} \\
%\Rightarrow\frac{\partial\tilde{\pi}}{\partial\tau} &= -\frac{2a_{1}T \tilde{\pi} }{3b_{1}} - \frac{\tilde{\pi}}{2}\left[\frac{1}{\tau} - 5\frac{1}{T} \frac{\partial T}{\partial\tau}\right] + \frac{8a_{1}T^{4}}{9\tau\pi^{0}} \\
\Rightarrow\frac{\partial\tilde{\pi}}{\partial\tau} &= -\frac{2a_{1}T_{0} \tilde{T} \tilde{\pi} }{3b_{1}} - \frac{\tilde{\pi}}{2}\left[\frac{1}{\tau} - 5\frac{1}{\tilde{T}} \frac{\partial \tilde{T}}{\partial\tau}\right] + \frac{8\tilde{T}^{4}}{9\tau b_{3}}. \\
\label{eq:58}
\end{aligned}
\end{equation}

From Eq.~(\ref{eq:57}) and Eq.~(\ref{eq:58}), the energy conservation equation and shear viscosity equation can be written as

\begin{equation}
\begin{aligned}
\frac{\partial \Tilde{T}}{\partial \tau} &= - \frac{\tilde{T}}{3\tau} + \frac{\epsilon_1}{\tilde{T}^{3}}\left(\frac{\tau_{0}}{\tau}\right)^{2a}\frac{1}{\tau} + \frac{b_{3}\tilde{\pi}\tilde{T}}{12 \tau},
\label{eq:59}
\end{aligned}
\end{equation}
\begin{equation}
\begin{aligned}
\frac{\partial\tilde{\pi}}{\partial\tau} &= -\frac{2a_{1}T_{0} \tilde{T} \tilde{\pi} }{3b_{1}} - \frac{\tilde{\pi}}{2}\left[\frac{1}{\tau} - 5\frac{1}{\tilde{T}} \frac{\partial \tilde{T}}{\partial\tau}\right] + \frac{8\tilde{T}^{4}}{9\tau b_{3}}. \\
\label{eq:60}
\end{aligned}
\end{equation}

We numerically solved the above second order (IS) approximation fluid differential equations, Eq.~(\ref{eq:59}) and Eq.~(\ref{eq:60}), using fixed initial conditions, and compared the second order (IS) results to those results obtained from ideal MHD and first order (NS) theory.
\begin{figure}[tbp!]
\includegraphics[width=0.85\linewidth]{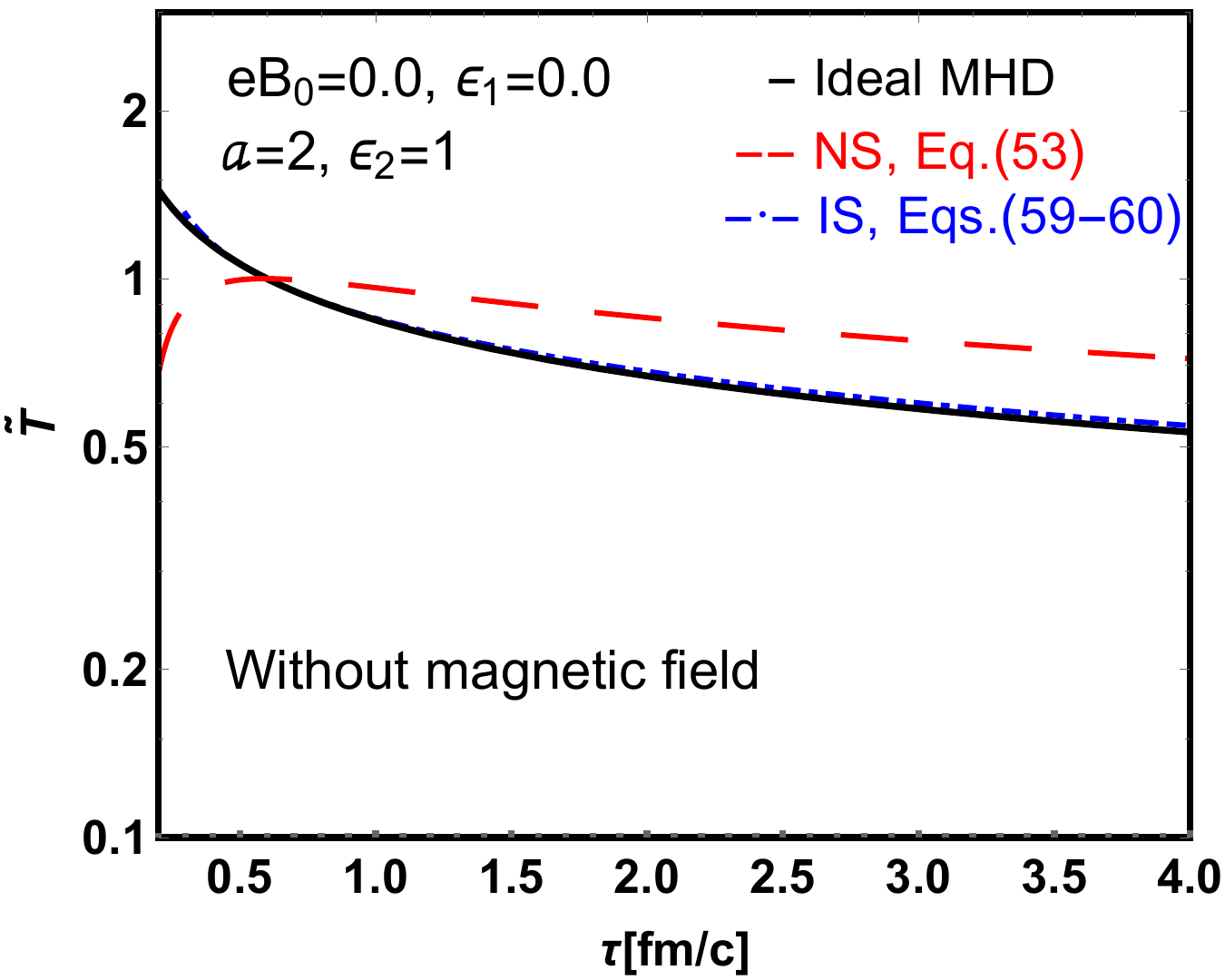} \\
\includegraphics[width=0.85\linewidth]{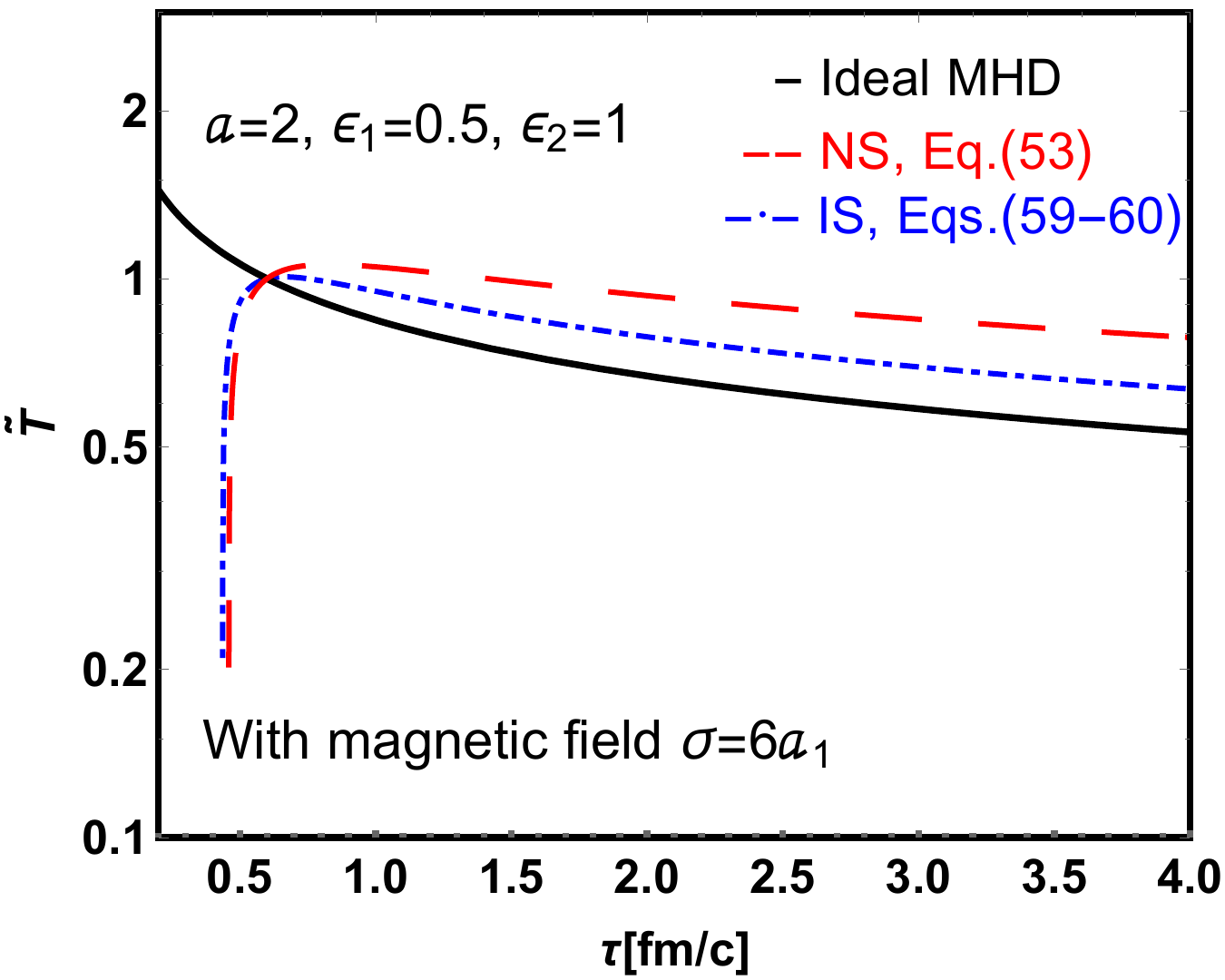}
\caption{(Color online) Evolution of normalized temperature $\widetilde{T}$ (Eqs.~(\ref{eq:53}, \ref{eq:59}, \ref{eq:60})) as a function of proper time $\tau$ for different initial magnetic field ($\epsilon_{1}$) and fixed shear viscosity value $\epsilon_{2}=1$, compared to the ideal MHD and first-order (NS) solution. The magnetic field decay parameter $a=2$. }
\label{f:is-1}
\end{figure}

In the upper panel of Fig.~\ref{f:is-1}, we present the normalized temperature $\tilde{T}$ as a function of proper time $\tau$ for ideal MHD, first order (NS) solution and second order (IS) with a zero magnetic field. The shear viscosity $\epsilon_{2}$ is $\epsilon_{2}=1$. The initial condition for $\pi_{0}$ is $\pi_{0}=0.1p_{0}$ where $b_{3}=0.1$. In this figure a comparison between the perfect fluid approximation, the first order theory, and the second order theory is clear. One finds that considering a non-zero shear viscosity makes the fluid cool down more slowly compared to the ideal MHD. Ones sees that there is a peak in $\tilde{T}$ in case of the first-order (NS) theory, and no peak in the second-order (IS) theory.

In the lower panel of Fig.~\ref{f:is-1}, the normalized temperature $\tilde{T}$ is plotted against proper time $\tau$ for three different scenarios: ideal MHD, first-order (NS) theory, and second-order (IS) theory, all with non-zero magnetic field and shear viscosity. The specific parameters used are: magnetic field decay parameter $a=2$, magnetic field parameter $\epsilon_{1}=0.5$, and the shear viscosity parameter $\epsilon_{2}=1$. Additionally, the initial condition for $\pi_{0}$ (a component related to shear stress) is set to $\pi_{0}=0.1p_{0}$, where $b_{3}=0.1$. The results indicate that the presence of a magnetic field leads to a slower cooling rate of the fluid compared to the case where only shear viscosity is considered. Furthermore, both the first-order and second-order theories exhibit a peak in $\tilde{T}$ at very early times in the system's evolution. The cooling rate of temperature in the second-order IS theory is slower than that in ideal MHD but faster than in the first-order NS theory.

\begin{figure}[tbp!]
\includegraphics[width=0.85\linewidth]{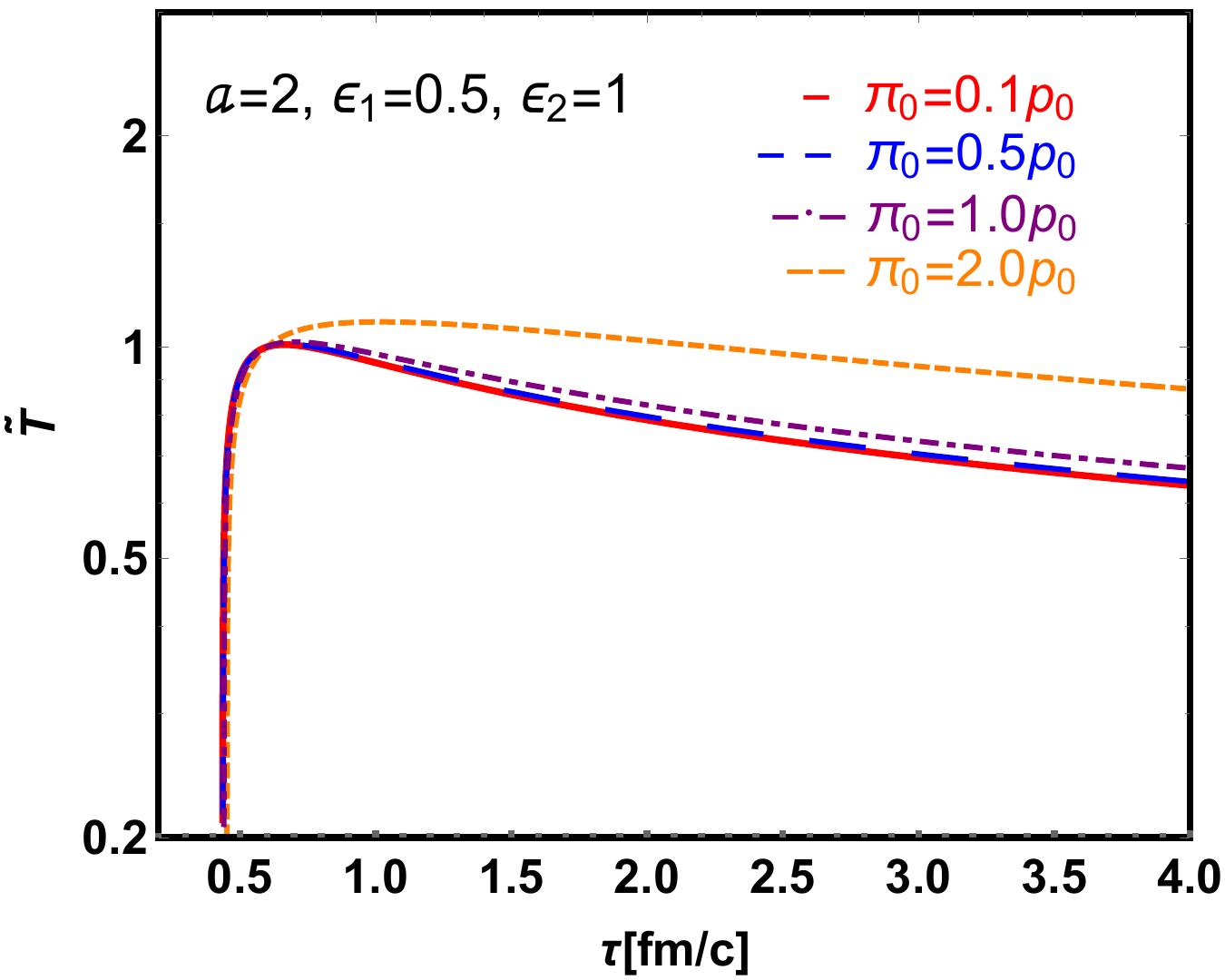}
\caption{(Color online) Evolution of normalized temperature $\widetilde{T}$ (\ref{eq:59}, \ref{eq:60})) from second-order (IS) theory as a function of proper time $\tau$ for different initial value of shear viscosity $\pi_{0}$ with fixed initial magnetic field $\epsilon_{1}=0.5$ and shear viscosity parameter $\epsilon_{2}=1$. The magnetic field decay parameter $a=2$.}
\label{f:is-2}
\end{figure}

In Fig.~\ref{f:is-2}, the normalized temperature $\tilde{T}$ is plotted as a function of proper time $\tau$ for the second-order (IS) theory, with a focus on the effect of different initial conditions for $\pi_{0}$. The magnetic field decay parameter $\epsilon_{1}$ is set to $0.5$ and the magnetic field decay parameter $a$ is 2. We compare four different initial conditions for $\pi_{0}$: $0.1p_{0}$, $0.5p_{0}$, $1p_{0}$ and $2p_{0}$, where $p_{0}$ is the initial pressure of the fluid. We find that a larger initial value of $\pi_{0}$ (or, a larger value of $b_{3}$) leads to a slower cooling rate of the fluid compared to smaller initial values of shear viscosity. This suggests that the initial shear stress in the fluid plays a significant role in determining its thermal evolution. Additionally, in all cases, there is a peak in $\tilde{T}$ observed at early times, which is a characteristic of viscous non-resistive magnetohydrodynamics. This peak arises due to the non-zero magnetic field heated up the fluid, which influence the fluid's cooling behavior.

%%%%%%%%%%%%%%%%%%%%%%%%%%%%%%%%%%%%%%%%%%%%%%%%%
\section{Conclusions}
\label{section-5}

Motivated by the exploration of strong magnetic fields and shear viscosity in relativistic heavy-ion collisions, we investigate the evolution of flow temperature in a 1+1 dimensional viscous, non-resistive magnetohydrodynamic flow with an EOS of $\varepsilon=3p$. Our idealized setup, focusing on one-dimensional, longitudinal boost-invariant flow with a transverse magnetic field and constant shear viscosity, yield various analytical solutions.

This work extends the Victor-Bjorken ideal MHD flow~\cite{Roy:2015kma,Pu:2016ayh} and Azwinndini-Bjorken dissipative flow~\cite{Muronga:2001zk,Muronga:2003ta} to scenarios incorporating both non-zero shear viscosity and a magnetic field. Specifically, the presented analytical solution reduces to Bjorken flow solution in the absence of both, to \textrm{Victor-Bjorken} type solution with a magnetic field but no shear viscosity, and to Azwinndini-Bjorken type solution with shear viscosity but no magnetic field. With both magnetic field and first-order (Navier-Stokes) shear viscosity, we obtain two new perturbative solutions and compared them with the numerical solutions. Additionally, we obtained numerical solutions for different initial shear viscosities $\pi_{0}$ in the second-order (Israel-Stewart) theory. Although simplified, our findings provide valuable understanding into the fluid dynamics of nucleus-nucleus collisions.

We further investigate scenarios with arbitrary shear viscosity and a small magnetic field evolving according to a power-law in proper time with exponent $a$. For initial magnetic field strengths ($B^{2}_{0}$) smaller than 6$a_{1}T^{4}_{0}$ and for proper time $\tau > 0.6$ fm/c, our analytical solution is stable. We observe that larger magnetic fields ($\epsilon_{1}$) with decay parameter $a>1$ result in fluid heating, manifesting as an early temperature peak whose magnitude depends on the magnetic field strength and shear viscosity. However, at late times, its temperature asymptotically decreases that is the same as in the Azwinndini-Bjorken flow.
We also consider the case where the shear viscosity and magnetic field are both small. The analytical solution in this case is consistented with the numerical results while the magnetic field is pretty small ($\epsilon_{1}<0.02$), but deviates to become 5\% greater than the numerical results while $\epsilon_{1}>0.05$.

Finally, we present the numerical solution in the second-order (Israel-Stewart) theory. We find that the presence of magnetic field and shear viscosity leads to a slower cooling rate of the fluid temperature compared to the case where only shear viscosity is considered. We also observe the initial shear stress $\pi_{0}$ in the longitudinal boost-invariant fluid plays a significant role in determining its temperature evolution.

\begin{acknowledgements}
We thank Shi Pu and Shen-Qin Feng for helpful discussions and comments. This work was supported by the National Natural Science Foundation of China (NSFC) under Grant Nos.~12305138. Duan She's research is funded by the Startup Research Fund of Henan Academy of Sciences (No. 231820058) and the 2024 Henan Province International Science and Technology Cooperation Projects (No. 242102521068).
\end{acknowledgements}

\bibliographystyle{unsrt}
\bibliography{clv3}

\end{document}